\documentclass[12pt,letterpaper]{article}

\usepackage[margin=1in]{geometry}
\usepackage{amsmath}
\usepackage{amssymb}
\usepackage{hyperref}
\usepackage{booktabs}
\usepackage{array}
\usepackage{longtable}
\usepackage{enumitem}
\usepackage{titlesec}
\usepackage{parskip}
\usepackage{microtype}
\usepackage{xcolor}
\usepackage{fancyhdr}
\usepackage{graphicx}
\usepackage[numbers]{natbib}  
\usepackage{booktabs}
\usepackage{colortbl}
\usepackage{xcolor}
\usepackage{rotating}
\usepackage{array}
\usepackage{multirow}
\usepackage[toc]{glossaries}

\definecolor{blue1}{RGB}{198,213,240}  
\definecolor{blue2}{RGB}{122,162,210}  
\definecolor{blue3}{RGB}{49,100,175}   
\definecolor{blue4}{RGB}{12,53,130}    

\usepackage{pifont}
\usepackage{bbding}
\newcommand*\ON[0]{\Checkmark}

\hypersetup{
    colorlinks=true,
    linkcolor=blue!70!black,
    urlcolor=blue!70!black,
    citecolor=blue!70!black
}

\makeglossaries
\newacronym{6d}{6D}{six-dimensional}
\newacronym{ags}{AGS}{Alternating Gradient Synchrotron}
\newacronym{ai}{AI}{artificial intelligence}
\newacronym{aiml}{AI/ML}{artificial intelligence/machine learning}
\newacronym{anl}{ANL}{Argonne National Laboratory}
\newacronym{bnl}{BNL}{Brookhaven National Laboratory}
\newacronym{cern}{CERN}{European Organization for Nuclear Research}
\newacronym{csns}{CSNS}{China Spallation Neutron Source}
\newacronym{dc}{DC}{direct current}
\newacronym{doe}{DOE}{U.S. Department of Energy}
\newacronym{dune}{DUNE}{Deep Underground Neutrino Experiment}
\newacronym{eic}{EIC}{Electron-Ion Collider}
\newacronym{ess}{ESS}{European Spallation Source}
\newacronym{fast}{FAST}{Fermilab Accelerator Science and Technology facility}
\newacronym{fcc}{FCC}{Future Circular Collider}
\newacronym{fcc-ee}{FCC-ee}{Future Circular Collider electron-positron collider}
\newacronym{fibre}{FIBRE}{\acrshort{fast}/\acrshort{iota} Bunch Rotation Experiment}
\newacronym{fnal}{FNAL}{Fermi National Accelerator Laboratory}
\newacronym{fsu}{FSU}{Florida State University}
\newacronym{gard}{GARD}{U.S. Department of Energy General Accelerator Research and Development program}
\newacronym{hpt}{HPT}{high-power target/targetry}
\newacronym{hts}{HTS}{high-temperature superconductor/superconducting}
\newacronym{ilc}{ILC}{International Linear Collider}
\newacronym{imcc}{IMCC}{International Muon Collider Collaboration}
\newacronym{iota}{IOTA}{Integrable Optics Test Accelerator}
\newacronym{ir}{IR}{interaction region}
\newacronym{isis}{ISIS}{ISIS Neutron and Muon Source}
\newacronym{jparc}{J-PARC}{Japan Proton Accelerator Research Complex}
\newacronym{jlab}{JLab}{Thomas Jefferson National Accelerator Facility/Jefferson Lab}
\newacronym{kek}{KEK}{High Energy Accelerator Research Organization, Japan}
\newacronym{lace}{LACE}{laser-assisted charge exchange}
\newacronym{lampf}{LAMPF}{Los Alamos Meson Physics Facility (renamed \acrshort{lansce} in 1995)}
\newacronym{lanl}{LANL}{Los Alamos National Laboratory}
\newacronym{lansce}{LANSCE}{Los Alamos Neutron Science Center}
\newacronym{lbnf}{LBNF}{Long-Baseline Neutrino Facility}
\newacronym{lbnl}{LBNL}{Lawrence Berkeley National Laboratory}
\newacronym{ldrd}{LDRD}{Laboratory Directed Research and Development}
\newacronym{lhc}{LHC}{Large Hadron Collider}
\newacronym{lts}{LTS}{low-temperature superconductor/superconducting}
\newacronym{map}{MAP}{Muon Accelerator Program}
\newacronym{mdi}{MDI}{machine-detector interface}
\newacronym{melody}{MELODY}{Muon station for sciEnce technoLOgy and inDustrY}
\newacronym{mice}{MICE}{Muon Ionization Cooling Experiment}
\newacronym{nc}{NC}{normal-conducting}
\newacronym{nhmfl}{NHMFL}{National High Magnetic Field Laboratory}
\newacronym{ornl}{ORNL}{Oak Ridge National Laboratory}
\newacronym{pipii}{PIP-II}{Proton Improvement Plan II}
\newacronym{psi}{PSI}{Paul Scherrer Institute}
\newacronym{rnd}{R\&D}{research and development}
\newacronym{radiate}{RaDIATE}{Radiation Damage In Accelerator Target Environments}
\newacronym{rcs}{RCS}{rapid-cycling synchrotron}
\newacronym{rebco}{REBCO}{rare-earth barium copper oxide}
\newacronym{rf}{RF}{radio frequency}
\newacronym{rla}{RLA}{recirculating linear accelerator}
\newacronym{rms}{RMS}{root mean square}
\newacronym{sc}{SC}{superconducting}
\newacronym{slac}{SLAC}{SLAC National Accelerator Laboratory}
\newacronym{sns}{SNS}{Spallation Neutron Source}
\newacronym{sps}{SPS}{Super Proton Synchrotron}
\newacronym{srf}{SRF}{superconducting radio frequency}
\newacronym{trl}{TRL}{technology readiness level}
\newacronym{usmcc}{USMCC}{U.S. Muon Collider Collaboration}
\newacronym{mcf}{$\mu$CF}{muon-catalyzed fusion}
\newacronym{mic}{$\mu$IC}{muon-ion collider}
\newacronym{mutristan}{$\mu$TRISTAN}{muon Transposable Ring Intersecting Storage Accelerators}
\newacronym{msr}{$\mu$SR}{muon spin rotation, relaxation, and resonance}
\newacronym{stfc}{STFC}{Science and Technology Facilities Council}
\newacronym{triumf}{TRIUMF}{TRI-University Meson Facility}
\newacronym{utk}{UTK}{University of Tennessee, Knoxville}

\titleformat{\section}{\large\bfseries}{\thesection}{1em}{}
\titleformat{\subsection}{\normalsize\bfseries}{\thesubsection}{1em}{}
\titleformat{\subsubsection}{\normalsize\itshape\bfseries}{\thesubsubsection}{1em}{}

\begin{document}


\vspace*{5mm}
\begin{center}
    {\LARGE\bfseries Report of the National Lab Accelerator \\[6pt]
    Study Group for a Muon Collider}\\[1em]
\end{center}

\vspace*{5mm}
\begin{figure}[h]
\centering
\includegraphics[width=0.85\textwidth]{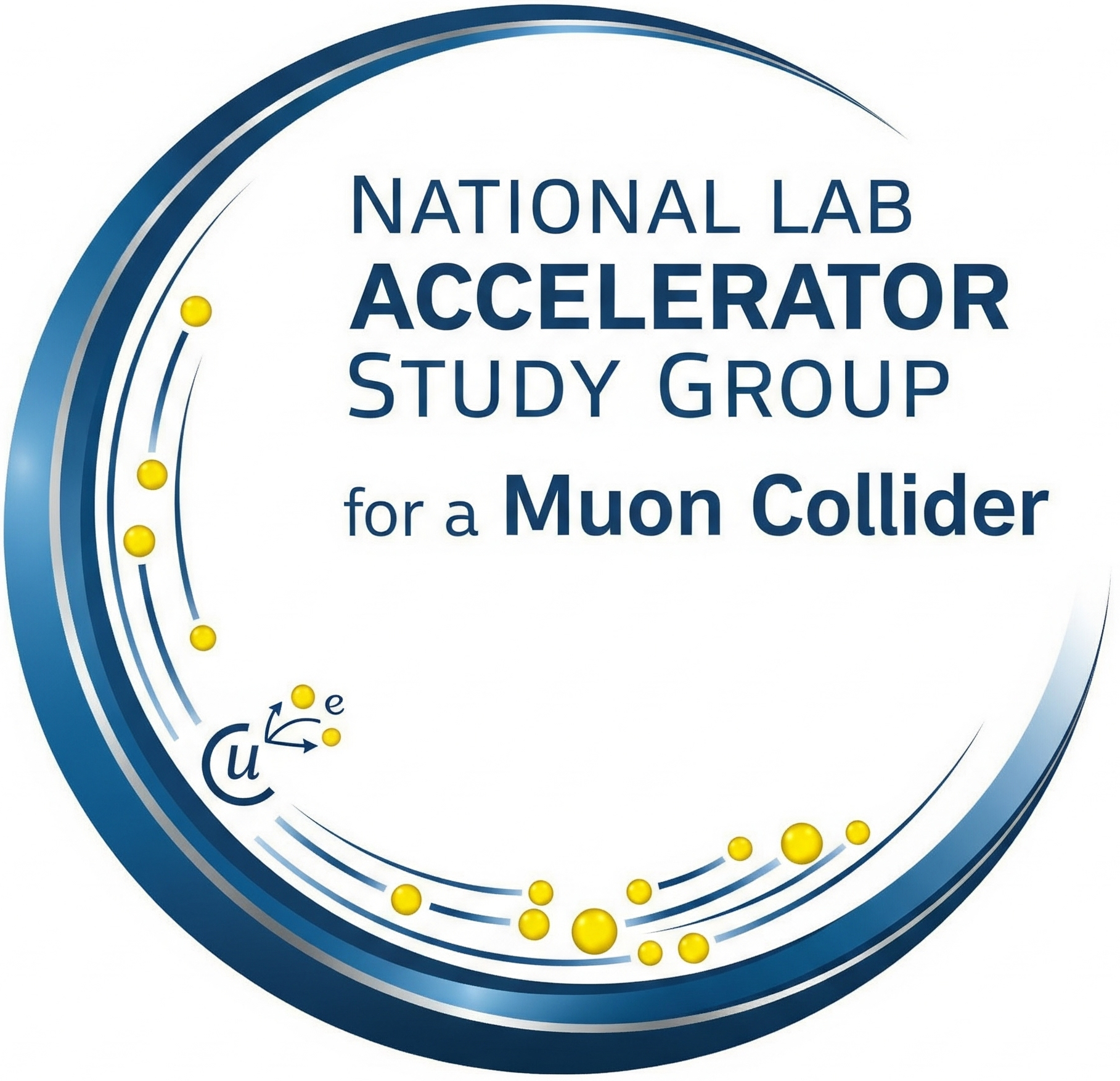}
\end{figure}

\vspace*{5mm}
\begin{center}    
{\large \today}\\   
\end{center}

\vspace{10mm}
{\bf Contributors:}  
S. Gourlay (\acrshort{fnal}, Chair),
C. Mitchell (\acrshort{lbnl}),
E. Nanni (\acrshort{slac}),
S. Peggs (\acrshort{bnl}),
F. Pilat (\acrshort{ornl}),
P. Piot (\acrshort{anl}),
R. Rimmer (\acrshort{jlab}),
S. Russell (\acrshort{lanl}).

\clearpage
~~
\vspace{2cm}
\paragraph{Abstract: } The National Laboratory Accelerator Study Group (NLASG) was established to assess the accelerator research and development required to advance a future high-energy Muon Collider, with particular emphasis on opportunities for coordinated contributions from U.S. national laboratories. This report examines the principal accelerator subsystems, their technical maturity, and the critical challenges that must be addressed before proceeding to a technical design. Topics include the proton driver, target and capture system, muon cooling, acceleration, collider ring, and machine-detector interface. The study identifies priority R\&D activities, evaluates opportunities to leverage existing laboratory capabilities and accelerator infrastructure, and outlines a staged approach to technology demonstrations and integrated system validation. Particular emphasis is placed on developing a coordinated national research strategy that complements international efforts and establishes a technically credible path toward a future Muon Collider.

\clearpage
\setcounter{tocdepth}{2}
\tableofcontents
\vspace*{20 mm}
\listoftables
\vspace*{20 mm}
\listoffigures

\clearpage

\section{Executive Summary}

A high-energy Muon Collider remains an  attractive option for extending the lepton energy frontier.
However, its realization presents an unprecedented series of accelerator challenges.

The National Lab Accelerator Study Group was established in August 2025 to provide a coordinated U.S. perspective on whether the technologies required for a 10~TeV Muon Collider can collectively reach the performance necessary for a practical facility.
This report assesses the status, challenges, and progress of design and technology development, after more than three decades of work and the recent efforts of the International Muon Collider Collaboration (\acrshort{imcc})~\cite{imcc}. 
Its goal is to outline the technical questions and the research and development (\acrshort{rnd}) infrastructure needed to answer them.
This is a preliminary step in determining whether a Muon Collider can ultimately be transformed from a compelling concept into a technically credible facility.

The short muon lifetime requires production, capture, cooling, acceleration, and collisions to occur very quickly. 
This demands extreme performance from accelerator subsystems. 
High-power targetry, six-dimensional cooling, high-gradient radio frequency (\acrshort{rf}) cavities in strong magnetic fields, high-field superconducting (\acrshort{sc}) magnets, rapid acceleration, machine protection, beam dynamics, detector backgrounds, and neutrino radiation must all achieve performance parameters that go beyond  or are at the limits of present capabilities. 
The requirements on these technologies are strongly coupled: they cannot be developed independently.

Studies have not identified any fundamental physical principle that precludes a Muon Collider. 
However, this does not demonstrate technical feasibility. 
Many important technologies remain at the conceptual, simulation, component-development, or early demonstration stage.
The complete sequence from muon production through cooling, acceleration and collisions has never been demonstrated as an integrated system. 
The principal question is therefore not whether individual concepts appear plausible, but whether a complete collider can be engineered with sufficient performance, reliability, efficiency, and cost effectiveness.

This challenge will require a new \acrshort{rnd} paradigm. 
Conventional accelerator development proceeds through incremental technical improvements followed by construction and operation of increasingly complex accelerators. 
A Muon Collider demands a more integrated approach in which simulation, component development, systems engineering, experimental validation, and collider design evolve together. 
Early experiments should not only demonstrate individual technologies, but also retire risks to the feasibility of the complete machine. 
Integrated test facilities and progressively more representative demonstrations will ultimately be required before a credible technical design can be established.

Although this report focuses on a 10~TeV center-of-mass collider with typical top-level parameters listed in Table~\ref{tab:baseline_10tev}, lower-energy facilities, perhaps including a 3 TeV and potentially a Higgs-factory-scale implementation, would provide important intermediate steps. 
Such intermediate facilities would not simply be scaled versions of a 10 TeV collider.
Rather, they would be elements of a staged development strategy in which scientific capability and technology validation proceed together.

The U.S. particle physics program is presently committed to completing and exploiting the High-Luminosity Large Hadron Collider (\acrshort{lhc})~\cite{BejarAlonso:2749422} and \acrshort{lbnf}/\acrshort{dune}~\cite{acciarri2016_short}.
Substantial dedicated funding for a Muon Collider \acrshort{rnd} program is therefore not imminent. 

The accelerator community must preserve critical expertise during the near-term period of limited funding, pursuing high-leverage \acrshort{rnd} that is synergistic with other accelerator programs, refining the collider concept, and identifying experiments and facilities that would be required for a future directed \acrshort{rnd} program.
The objective of the near-term phase should not be to prematurely establish a construction timeline, but to determine -- through focused experiments, integrated demonstrations, and rigorous systems analysis -- whether the required technologies can collectively reach the performance necessary for a practical collider.

The long-term path to a Muon Collider must be viewed as a technically limited development program, rather than as an extension of existing technology on a conventional project schedule. 
It will take time to establish the technical basis for a construction decision, even with sustained and adequately funded \acrshort{rnd}.

The diverse muon accelerators that are running or are under discussion (see Table~\ref{T:OTHER}) have common technical and physical challenges that distinguish them from more conventional electron and ion accelerators.  
In all cases muons must be created, cooled, collected, and delivered, all while respecting the relatively short muon lifetime.  
In some cases, like muon-catalyzed fusion, these common challenges can be met by common accelerator solutions.  
In other cases multiple solutions are necessary for different muon accelerators.  
A balanced Muon Accelerator \acrshort{rnd} Plan would deliver a better understanding of these commonalities and differences, strengthening U.S. capabilities across the entire evolving field of muon accelerators and their applications.

\begin{table}[h]
\centering
\caption{Typical top-level parameters for a 10~TeV Muon Collider~\cite{consolidatedParameters}.}
\label{tab:baseline_10tev}
\begin{tabular}{l c c c}
\toprule
{\bf Parameter} 		& {\bf Symbol} 		& {\bf Unit} 	& {\bf Value} \\
\midrule
Center-of-mass energy & $E_{\mathrm{cm}}$ & TeV & 10 \\
Target integrated luminosity & $\int \mathcal{L}_{\mathrm{target}}$ & ab$^{-1}$ & 10 \\
Estimated luminosity & $\mathcal{L}_{\mathrm{estimated}}$ & $10^{34}\,\mathrm{cm^{-2}\,s^{-1}}$ & 18 \\
Collider circumference & $C_{\mathrm{coll}}$ & km & 11.4 \\
Collider arc peak field & $B_{\mathrm{arc}}$ & T & 14 \\
Luminosity lifetime & $N_{\mathrm{turn}}$ & turns & 1363 \\
\midrule
Muons per bunch & $N$ & $10^{12}$ & 1.8 \\
Repetition rate & $f_r$ & Hz & 5 \\
Beam power & $P_{\mathrm{coll}}$ & MW &14.4 \\
RMS longitudinal emittance & $\epsilon_{\parallel}$ & eV$\cdot$s & 0.025 \\
Norm. RMS transverse emittance & $\epsilon_{\perp}$ & $\mu$m & 25 \\
\midrule
Interaction point bunch length & $\sigma_z$ & mm & 1.5 \\
Interaction point beta function & $\beta$ & mm & 1.5 \\
Interaction point beam size & $\sigma$ & $\mu$m & 0.9 \\
\midrule
Protons on target per bunch & $N_p$ & $10^{14}$ & 5 \\
Proton energy on target & $E_p$ & GeV & 5 \\
\bottomrule
\end{tabular}
\end{table}

\clearpage
\section{Responses to the Four Charge Questions}

\subsection{Identify the key accelerator challenges that must be addressed to enable the realization of a 10\,TeV Muon Collider~\label{sec:detail_on_challenges}}

The key challenges are addressed in detail in the following sub-sections.  
In brief, they are:

\begin{itemize}
\item {\bf Proton driver.} 
Transport and compress multi-MW beams with ultra-short bunch lengths, subject to intense space charge.
\item {\bf Target and capture.} 
Mitigate the extreme thermal and mechanical stresses on the proton target, maintaining efficient hadron capture. 
\item {\bf Beam cooling.} 
Demonstrate a complete integrated cooling system that rapidly shrinks the six-dimensional (\acrshort{6d}) muon phase space volume.
\item {\bf Rapid acceleration.} 
Accelerate intense bunches within the short muon lifetime while preserving all three emittances.
\item {\bf Collider ring.} 
Provide sufficient momentum aperture and luminosity for beams with a large energy spread, in the presence of beam-beam and wakefield effects, while mitigating the radiation load within the collider and off-site.
\item {\bf Machine--Detector Interface.} 
Establish a lattice that delivers a small transverse beam size at the interaction point while controlling the beam halo.
Meet engineering challenges for the detector nozzles and the central beam chamber.
\end{itemize}

\subsubsection{Proton driver}

The proton driver is a foundational component of a Muon Collider, positioned upstream of all muon systems. 
The parameter goals listed in Table~\ref{T:DRIVER} must be met, with ultra-short bunches of high intensity that are well beyond the capability of existing and near-term megawatt-class facilities (such as the Spallation Neutron Source (\acrshort{sns}) and the Proton Improvement Plan-II (\acrshort{pipii})).
The reference design consists of a high-power H$^-$ linac feeding an accumulator ring and a compressor ring, which together generate and compress high-intensity proton pulses for delivery to the target. 
Key design choices must reflect tradeoffs among beam power, repetition rate, and operational constraints.

\begin{table}[h!]
\centering
\caption[Parameter goals for the proton driver]
{Parameter goals for the proton driver.}
\begin{tabular}{lccl}
\toprule
\textbf{Parameter} & \textbf{Value} & \textbf{Unit} & \textbf{Notes} \\
\midrule
Beam power           & 2--4       & MW      & Delivered to the production target \\
Energy range                  & 5--12      & GeV     & Broad flat muon-yield region  \\
Energy optimum       & $\sim$8    & GeV     & \acrshort{bnl} \& \acrshort{fnal} compatible \\
RMS bunch length                   & 1--3       & ns      & Ultra-short \\
Protons per bunch              & $\sim$10$^{14}$ &   & High-intensity pulses \\
Repetition rate                & 5--10      & Hz      & Baseline $\sim$10\,Hz \\
\bottomrule
\end{tabular}
\label{T:DRIVER}
\end{table}

Charge-exchange injection at multi-megawatt levels likely requires laser-assisted techniques, as conventional foil stripping becomes impractical. 
Extreme space-charge effects during bunch compression challenge the achievement of nanosecond-scale bunches, motivating studies of advanced \acrshort{rf} systems and mitigation methods. Examples include rapid ``snap" longitudinal compression schemes, with proposed \acrshort{rnd} programs at \acrshort{sns} and/or Fermilab Accelerator Science and Technology/Integrable Optics Test Accelerator (\acrshort{fast}/\acrshort{iota}).
Experimental validation and continued simulation efforts are essential to reduce beam dynamics uncertainties and risks. 
Additional design considerations include phase-slip factor optimization~\cite{Hwang_2001} and the selective use of bunch combiners.

\subsubsection{Target and capture}

The high-power target (\acrshort{hpt}) and capture system must operate under beam conditions beyond those of existing facilities. 
Top-level target system goals are listed in Table~\ref{T:TARGET}.
The target must efficiently produce pions while withstanding extreme energy deposition, radiation damage, and thermal and mechanical stresses. 
The target operates with multi-megawatt proton beams, nanosecond-scale bunch lengths, and small spot sizes.
This imposes challenging demands on materials, cooling, and structural design. 
Development relies on a combination of simulations, irradiation studies, and scaled experiments because no facility can fully replicate these conditions.

\begin{table}[h!]
\centering
\caption[Target system]
{Parameter goals for the target system.}
\begin{tabular}{lccl}
\toprule
{\bf Parameter} 		& {\bf Value} & {\bf Unit} & {\bf Notes} \\
\midrule
Muon yield  		& $>$10     	& \%		& Muons generated per incident proton \\
Operational lifetime	& $\sim$1    	& year	& Per target assembly \\
Solid target power	& $\sim$2--4 	& MW	& With advanced cooling \\
\bottomrule
\end{tabular}
\label{T:TARGET}
\end{table}

Key design goals include high muon yield, efficient capture, effective heat removal, low activation, remote handling capability, and operational lifetimes of about one year. 
Cross-cutting \acrshort{rnd} efforts, including radiation damage studies and integrated simulations, are essential to evaluate performance and guide design choices. 
The program aims to converge on a demonstrator target design, with priorities focused on comparing target concepts, validating material behavior under extreme conditions, and developing a realistic, testable system.

\subsubsection{Beam cooling}

Beam cooling is one of the most demanding subsystems. 
After capture, an advanced and technically complex cooling system is required to rapidly reduce the muon bunch's \acrshort{6d} phase space volume within the short muon lifetime.
Cooling methods under development include ionization cooling, which reduces transverse emittance by passing the muon beam through low-$Z$ absorbers, with \acrshort{rf} cavities that restore longitudinal momentum.
Alternative cooling concepts -- such as cooling rings and helical channels -- also continue to be investigated. 

Modeling indicates that ionization cooling can achieve the required performance, although several technical challenges remain, including:

\begin{itemize}
\item operating \acrshort{rf} cavities in strong magnetic fields,
\item integrating high-field solenoids with absorbers and \acrshort{rf} systems,
\item controlling beam loading effects,
\item understanding achievable tolerances, and
\item precise matching between cooling stages.
\end{itemize}

Experimental results from the Muon Ionization Cooling Experiment (\acrshort{mice})~\cite{mice_cooling_2020}, muCool~\cite{Belosevic2019muCool} and elsewhere provide important partial validation of ionization cooling, but the full demonstration of an integrated cooling channel is still necessary.
Key \acrshort{rnd} priorities include:

\begin{itemize}
\item high-field magnets,
\item \acrshort{rf} power generation and distribution,
\item system integration,
\item advanced modeling, and 
\item development of a realistic demonstrator system.
\end{itemize}

\subsubsection{Rapid acceleration}

The short muon lifetime requires rapid acceleration from a few~MeV to many~TeV, while preserving the beam emittances and high bunch charges.
A staged approach uses different technologies at different energies, with typical parameters listed in Table~\ref{T:RMA}.

\begin{table}[h!]
\centering
\caption[Rapid muon acceleration]
{Parameter goals for rapid muon acceleration.}
\setlength{\tabcolsep}{4pt}
\begin{tabular}{lccl}
\toprule
{\bf Parameter} 				& {\bf Value} 			& {\bf Unit} 	& {\bf Notes} \\
\midrule
Muon kinetic energy after cooling			& $\sim$3--4   			& MeV   		& Input to acceleration chain \\
\acrshort{srf} linac output energy         & $\sim$3     			& GeV   		& Hand-off from linac to \acrshort{rla} \\
RLA output energy               & $\sim$60     			& GeV   		& \\
Maximum synchrotron energy		& $\sim 5$   			& TeV   		& After pulsed synchrotrons \\
Bunch population after cooling	& $\sim 3\times10^{12}$ &  			    &  \\
\hspace{32mm} at the collider	& $\sim 2\times10^{12}$ &  			    &  \\
Muon decay before high-frequency \acrshort{rf}	& $>$30        			& \%    		& During early acceleration \\
\acrshort{srf} cavity frequency			& $ \sim1.3$         			& GHz   		& 325 \& 650\,MHz also needed \\
\bottomrule
\end{tabular}
\label{T:RMA}
\end{table}
{\bf A linac} is used for early acceleration after initial cooling, because low-energy beams cannot be recirculated. 
This early stage is particularly difficult due to low \acrshort{rf} frequencies, limited gradients, and large bunch sizes, all of which can lead to significant decay losses.
Longitudinal emittance is a critical parameter.
{\bf Recirculating Linear Accelerators} (\acrshort{rla}s) are used at energies up to about 60~GeV, using superconducting \acrshort{rf} (\acrshort{srf}) systems with evolving \acrshort{rf} frequencies and gradients to balance efficiency and beam stability. 
{\bf Pulsed synchrotrons} become more efficient at energies above about 60~GeV, although their feasibility depends on projected advances in magnet technology. 
{\bf Hybrid designs} combine superconducting and pulsed magnets to offer a path to multi-TeV energies, at the cost of introducing complexity in beam dynamics and system design. 

Detailed simulations of these acceleration stages are needed, including component design, beam dynamics, wakefields, and matching. 
Key challenges include rapid magnet ramping, high-power \acrshort{rf} systems, and managing beam loading and longitudinal stability. 
Continued progress depends on advanced simulations and focused \acrshort{rnd}, particularly in pulsed magnets, \acrshort{rf}, and system integration.

\subsubsection{Collider ring}

The collider ring must control the beam near the interaction point to provide the desired luminosity in the presence of beam-beam interaction effects. 
The collider ring lattice, which integrates low-beta interaction regions, matching sections, and strongly focusing arcs, faces challenges from the large energy spread at injection and from strong chromatic effects. 
Top-level goal parameters are listed in Table~\ref{T:COLLIDER}.

\begin{table}[h!]
\centering
\caption[Collider ring]
{Parameter goals for the collider ring.}
\begin{tabular}{lccl}
\toprule
\textbf{Parameter} & \textbf{Value} & \textbf{Unit} & \textbf{Notes} \\
\midrule
Center-of-mass energy range         	& 3--10    				& TeV	&  \\
\acrshort{rms} beam energy spread            	& 0.1          			& \%    	& At injection \\
Beam--beam parameter, $\xi$                	& 0.078        			&     	& High bunch intensity \\
Beam-induced power deposition  	& $\sim$500 			& W/m   	& Dominates cold mass heat load \\
\bottomrule
\end{tabular}
\label{T:COLLIDER}
\end{table}
Sextupoles are effective in a local chromaticity correction scheme, but leave higher-order effects that limit energy acceptance and dynamic aperture, particularly at higher energies. 
Beam--beam interactions and impedance effects appear to be manageable with feedback systems and careful design, although further integrated simulations are needed to validate luminosity goals.

Advanced magnet designs must accommodate high fields, large apertures, and substantial internal shielding to handle intense decay-induced heat loads. 
A technology path from niobium-titanium magnets to niobium-tin to high-temperature superconductors (\acrshort{hts}) such as \acrshort{rebco} is envisioned.

Off-site neutrino radiation remains a key external constraint.
Mitigation strategies show promising reductions to acceptable levels within the range of existing facilities, at the expense of adding lattice design complexity.
For example, the VIPER scheme~\cite{Curco:2024} introduces vertical beam orbit modulation and a modified site layout.
Preliminary site and civil engineering studies at \acrshort{cern} and \acrshort{fnal} support the feasibility of these mitigation schemes, although more design work is required.

\subsubsection{Machine--Detector Interface}

The Machine--Detector Interface (\acrshort{mdi}) requires a tightly coupled optimization of the interaction region, detector nozzles, and overall detector design in order to balance competing demands. 
These include physics performance (minimizing beam-induced backgrounds), machine requirements, engineering and integration constraints, and radiation effects such as heat deposition and long-term material damage. 
The need for a very small $\beta^*$ at the interaction point impacts the lattice design, driving stringent requirements on momentum acceptance, magnet alignment, aperture size, and achievable field strength.
The beam halo needs careful control, potentially using a dedicated halo cleaning system. 

Significant engineering challenges arise in the design, assembly, and alignment of the detector nozzles, central beam chamber, and forward instrumentation such as luminosity monitors. 
Successful implementation requires early and comprehensive \acrshort{mdi} integration studies to define spatial constraints and interfaces among detector systems, accelerator components, and infrastructure.
This includes magnets, vacuum systems, cryogenics, and beam instrumentation.

\clearpage
\subsection{Evaluate the \acrshort{rnd} steps, demonstrations, simulation and software tools, and facility capabilities required to address these challenges~\label{sec:detail_on_steps}}

The \acrshort{rnd} strategy forms a distributed hierarchical program that enables national lab participation, with specialized demonstrators (e.g., targetry) contributing to a coordinated and flexible program. \acrshort{rnd} should progress from many small, targeted experiments to fewer integrated system-level demonstrations, perhaps including a lower energy unified intermediate facility. Small-scale demonstrators enable rapid learning and provide critical data to benchmark simulations. They build confidence in the validity of performance extrapolations and enable technical down-selections, while reducing risk and avoiding premature single-path commitment. 

The choice of technology for a \acrshort{fnal}-based proton driver should be considered within this framework, requiring open discussion of medium-term funding realities. Intermediate-stage high-energy physics results would strengthen the case for continued investment. Complementary advances in accelerator theory and simulation -- particularly in accelerator optics and dynamics -- are essential to guide development and technology choices.

Assigning a Technology Readiness Level (\acrshort{trl}) to technology \acrshort{rnd} provides a useful summary measure of its maturity, but by itself it is insufficient to assess the potential impact or timescale of a near-term \acrshort{rnd} investment.
Nonetheless, we use it here as a convenient evaluation of the key accelerator challenges.
Table~\ref{T:TRL} summarizes the definitions of levels 1 through~9.

\begin{table}[h!]
\centering
\caption[Summary definition of Technical Readiness Levels]{Summary definition of Technical Readiness Levels.}
\begin{tabular}{cl}
\toprule
{\bf \acrshort{trl}} 		& {\bf Definition} \\
\midrule
1   & Basic principles observed and reported    \\
2   & Technology concept and/or application formulated.  \\
3   & Analytical \& experimental proof of concept for critical functions \& characteristics.  \\
4   & Component and/or breadboard validation in a laboratory environment.\\
5   & Component and/or breadboard validation in a relevant environment. \\
6   & System or subsystem model/prototype demonstrated in a relevant environment. \\
7   & System prototype demonstrated in an operational environment. \\
8   & Actual system completed and qualified through testing and demonstration. \\
9   & Actual system proven through successful mission operations. \\
\bottomrule
\end{tabular}
\label{T:TRL}
\end{table}

In many cases ongoing generic \acrshort{rnd} programs, advances driven by industry, and planned upgrades to accelerator facilities will continue to mature key technologies independent of a dedicated Muon Collider program. These parallel efforts can significantly accelerate technology readiness, reduce development risk, and shorten the time required to achieve the milestones necessary for a Technical Design Report.

\subsubsection{Proton driver}

\paragraph{\acrshort{rnd} Priorities:}
Develop multi-MW proton driver capability beyond existing MW-class machines, including H$^-$ linacs, high-intensity accumulator and compressor rings, laser-assisted charge-exchange injection (\acrshort{lace}), bunch compression under extreme space charge, high-power \acrshort{rf} systems, and integrated end-to-end simulations to validate beam dynamics and target delivery. \acrshort{trl} assessments are shown in Table~\ref{tab:Proton Driver}.

\begin{table}[htbp]
\centering
\caption[Technical Readiness Levels -- proton driver]{Technical Readiness Levels -- proton driver.}
\label{tab:Proton Driver}
\begin{tabular}{p{70 mm} c p{73 mm}}
\toprule
\textbf{Key enabling \acrshort{rnd}} & \textbf{\acrshort{trl}} & \textbf{Comment}  \\
\midrule
Demonstrate \acrshort{lace} H$^{-}$ injection at high intensity. & 3 &
Actively pursued at \acrshort{sns}~\cite{PhysRevAccelBeams.26.043501}.  \\
\addlinespace
Validate bunch compression under extreme space charge. & 2 &
Could be tested in \acrshort{bnl}'s Alternating Gradient Synchrotron (\acrshort{ags}), \acrshort{sns} or \acrshort{iota}.  \\
\addlinespace
Develop integrated, end-to-end simulations. & 6 &
Ensure self-consistent analysis of physics and engineering using multiple codes.  \\
\bottomrule
\end{tabular}
\end{table}

\paragraph{Discussion:}
The proton driver is a foundational element of a Muon Collider. 
While not the dominant contributor to overall cost or technical risk, it is upstream of all muon systems and must be early in achieving a high \acrshort{trl}. 
Proton driver technology is closer to existing accelerator capabilities than downstream muon systems, so it provides an early opportunity to offload risk while enabling multiple potential staging options for high-energy physics.

The proton driver must significantly extend the performance of existing megawatt-class facilities such as \acrshort{sns} and Japan Proton Accelerator Research Complex (\acrshort{jparc}). Key parameter goals for beam power, proton energies, and ultra-short bunch lengths, et cetera, are listed in Table~\ref{T:DRIVER}.  
The reference architecture consists of a high-power H$^-$ linac, followed by an accumulator ring and a compressor ring. 
The linac accelerates protons to a low energy, the accumulator collects and organizes high-intensity pulses, and the compressor performs aggressive longitudinal phase-space rotation to compress bunch lengths from microseconds to nanoseconds. 
In multi-bunch configurations, a downstream combiner synchronizes arrival at the target.

Repetition rates of 5--10\,Hz are favored, with $\sim$10~Hz providing a practical balance. 
Proton energies near 8~GeV are attractive due to relatively flat muon yield and alignment with \acrshort{fnal} infrastructure. 
A central technical challenge is charge-exchange injection: conventional foil stripping becomes impractical at multi-megawatt intensities, making \acrshort{lace} injection a necessary solution. 
Significant progress in laser power and stripping efficiency has been made with a \acrshort{lace} demonstration~\cite{PhysRevAccelBeams.26.043501}. 
\acrshort{ornl} plans to install a \acrshort{lace} injection system at \acrshort{sns} in the next five years.

The accumulator and compressor rings must contend with extreme space-charge effects, particularly during bunch compression. 
Techniques under study include high-voltage \acrshort{rf} systems, resonant schemes, and impedance manipulation (e.g., inductive inserts). 
Proposed experiments, such as the \acrshort{fast}/\acrshort{iota} Bunch Rotation Experiment (\acrshort{fibre})~\cite{simons2025iotaexperimentprotonpulse}, aim to demonstrate bunch rotation under realistic space-charge conditions. 
International efforts, particularly at \acrshort{cern}, share similar linac approaches but explore alternative ring designs.

\subsubsection{Target and capture}

\paragraph{\acrshort{rnd} Priorities:}
Advanced target concepts capable of surviving multi-MW, nanosecond proton pulses, including solid, liquid-metal, and fluidized-particle targets. Key needs include radiation-damage studies, thermal shock validation, high-temperature material testing, remote handling, capture solenoid integration, and a realistic target demonstrator with clear performance criteria. 
\acrshort{trl} assessments are shown in Table~\ref{T:Target and Capture}. 

\begin{table}[h]
\centering
\caption[Technical Readiness Levels -- target and capture]{Technical Readiness Levels -- target and capture.}
\label{tab:technology_assessment}
\begin{tabular}{p{70 mm}c p{73 mm}}
\toprule
\textbf{Key enabling \acrshort{rnd}} & \textbf{\acrshort{trl}} & \textbf{Comment} \\
\midrule
Analyze and compare solid, liquid metal, and fluidized solid target concepts. & -- & Some support through the \acrshort{gard} and \acrshort{radiate} collaborations. 	 \\
\addlinespace
\;\;\; Solid target. & 5 &
1 MW achieved, 5~MW expected at \acrshort{ess}. 
Low power designs for current projects. \\
\addlinespace
\;\;\; Liquid metal target. & 3 &
Up to 2 MW at \acrshort{sns}.
Options being explored. Beam windows are an issue.\\
\addlinespace
\;\;\; Fluidized solid target. & 2 &
At an early stage of a long \acrshort{rnd} cycle.\\
\addlinespace
Validate material properties under extreme radiation and high temperature. & -- &\\
\addlinespace
\;\;\; Develop a realistic target concept. & 1 &
No directed \acrshort{rnd} effort.  \\
\addlinespace
\;\;\; Novel beam-intercepting materials. & 3 &
Conventional materials are at their limit. 
Early \acrshort{rnd} stage.\\
\addlinespace
Large-bore, high-field solenoids. & 3 &
Fundamental \acrshort{rnd} and small-scale demos (\acrshort{fsu}/MagLab and U.S. Magnet Development Program~\cite{usmdp}). 
Some large-scale development led by industry will be relevant. \\
\bottomrule
\end{tabular}
\label{T:Target and Capture}
\end{table}

\paragraph{Discussion:}
The realization of a Muon Collider depends critically on a high-power target system capable of withstanding beam conditions beyond any existing facility. 
Targets must operate with multi-megawatt proton beams, nanosecond-scale bunches, and small beam spot sizes, imposing unprecedented demands on materials, cooling, and structural integrity. 

In a reference design, the target must operate within a 20~T magnetic field where muons, pions, kaons and other secondary particles are produced. 
Other magnets range in strength from 20~T down to 1.5~T at a point 19~m from the target, with bore diameters up to 1.4~m. 
Parameter goals are listed in Table~\ref{T:TARGET} for muon yield efficiency, maximum capture efficiency, extreme heat load management, minimum activation, and operational lifetime.

Three primary target approaches are under active investigation:

{\bf Solid targets (e.g., graphite or tungsten)} benefit from operational simplicity and experience, with potential performance up to $\sim$2--4\,MW using advanced cooling. However, thermal shock and stress-wave damage remain critical limitations.

{\bf Liquid metal targets (e.g., mercury, lead, or lead-bismuth)} avoid solid radiation damage and enable external heat removal, but introduce challenges in containment, cavitation, magnetohydrodynamics, and window survivability.
The MERIT experiment~\cite{McDonald:2009zz} showed that a free liquid mercury jet can survive intense beam pulses in a strong magnetic field. 

{\bf Fluidized tungsten targets} offer strong thermal shock resistance and efficient heat removal by circulating granular material. Early experiments are promising, but erosion and system reliability remain open issues.

Radiation damage and material performance studies are coordinated through collaborations such as \acrshort{radiate}~\cite{RaDIATE} and the \acrshort{doe} General Accelerator \acrshort{rnd} (\acrshort{gard}) program. The \acrshort{rnd} strategy combines simulation (e.g., \href{https://fluka.cern/}{FLUKA}, \href{https://github.com/ansys}{ANSYS}) with targeted experiments and synergies with programs such as \acrshort{lbnf}~\cite{acciarri2016_short} and Mu2e~\cite{mu2e}.

\subsubsection{Beam cooling}

\paragraph{\acrshort{rnd} Priorities:}
Large-aperture high-field magnets, high-gradient \acrshort{rf} cavities operating in strong magnetic fields, beam-loading mitigation, absorber and cavity material validation, stage-to-stage matching, and integrated cooling-channel demonstrations. 
\acrshort{trl} assessments are shown in Table~\ref{T:Cooling}.

\paragraph{Discussion:}
Muon Collider performance depends critically on the ability to capture intense muon beams and reduce their \acrshort{6d} phase space within the limited muon lifetime. 
A coherent and credible approach to muon production and rectilinear cooling has recently emerged, with promising performance that is supported by detailed beam physics studies, but which requires experimental verification.

The muon front end captures and conditions beam from the target, using a tapered solenoidal channel. 
A solenoid chicane provides momentum filtering without dipole magnets, exploiting vertical dispersion to remove unwanted high-momentum particles. 
A beryllium absorber further removes low-momentum protons prior to bunch formation. 
A Neuffer-style buncher and phase rotator~\cite{10.1063/1.1818445} then adiabatically bunches the beam in \acrshort{rf} cavities and rotates it in longitudinal phase space.

Ionization cooling forms a core mechanism for reducing the transverse emittances. 
Momentum is lost in all directions as the beam passes through low-$Z$ absorbers, while \acrshort{rf} cavities restore only the longitudinal component.
This results in net transverse cooling. 
The \acrshort{mice} experiment demonstrated these effects on a short cooling channel section~\cite{mice_cooling_2020}; however, the later stages of \acrshort{rf} re-acceleration were not realized. 
The demonstration of a realistic integrated cooling channel section with \acrshort{rf} and beam remains an essential technical goal.

The rectilinear cooling channel employs combined-function dipole--solenoid magnets with \acrshort{rf} cavities embedded directly {\it between} the magnet cryostats, creating a dense lattice capable of strong focusing and efficient cooling. Managing the large  forces between closely spaced solenoid coils with opposite polarity is a significant engineering challenge.
Overall performance is governed by a ``leaky funnel'' behavior, motivating the use of multiple, progressively stronger, stages. 
\acrshort{rf} cavity operation in strong magnetic fields remains a key risk, though mitigation strategies appear to offer breakthrough possibilities. 
These include high-pressure gas-filled cavities~\cite{osti_1221361}, modified copper--beryllium surfaces~\cite{PhysRevAccelBeams.23.072001}, and cryogenic operation. 
Alternative concepts continue to be explored.

\begin{table}[t]
\centering
\caption[Technical Readiness Levels -- cooling and beam preparation]{Technical Readiness Levels -- cooling and beam preparation.}
\label{tab:technology_assessment}
\begin{tabular}{p{70 mm}c p{73 mm}}
\toprule
\textbf{Key enabling \acrshort{rnd}} & \textbf{\acrshort{trl}} & \textbf{Comment} \\
\midrule
Large-acceptance, high-gradient \acrshort{rf} cavities capable of operating in high magnetic fields. & 4 &
 Beryllium pillbox cavity tests at \acrshort{fnal} showed promising results. Alternative materials and engineered designs are needed. \\
\addlinespace
Model the dependence of \acrshort{rf} breakdown in high magnetic fields on material properties, to understand frequency scaling and breakdown dynamics. & 3 &
Fundamental \acrshort{rnd} is ongoing, supported by \acrshort{gard}. \\
\addlinespace
Validate material and cavity performance under high \acrshort{rf} and magnetic fields, including at cryogenic temperatures and with beam irradiation. & 3 &
Fundamental \acrshort{rnd} is ongoing, supported by \acrshort{gard}, to understand frequency scaling and test cryogenic operation in high magnetic field. \\
\addlinespace
Model cooling channel intensity-dependent dynamics, including space charge and wakefield effects. & 3 &
 Wakefield effects in closed cavities need to be included, especially after the bunch merge. \\
\addlinespace
Affordable high peak-power short-pulse \acrshort{rf} sources, and \acrshort{rf} distribution. & 3 &
Tens of MW peak power. \\
\addlinespace
Develop a realistic ionization cooling channel concept for a demonstration, integrating \acrshort{rf}, absorbers and magnets. & 2 & At least one practical module is needed to inform potential demonstrators. \\
\addlinespace
Demonstrate \acrshort{6d} ionization cooling with realistic beam, \acrshort{rf}, absorbers, and high-field solenoids.  & 2 &
A longer-term but essential demonstration goal. \\
\addlinespace
Cooling channel diagnostices and upstream/downstream emittance measurements.  & 3 &
Integrated into cooling channel demonstrator. \\
\bottomrule
\end{tabular}
\label{T:Cooling}
\end{table}

\subsubsection{Rapid acceleration}

\paragraph{\acrshort{rnd} Priorities:}
Develop acceleration systems capable of taking muons from a few MeV to multi-TeV energies before excessive decay losses occur. 
Low-frequency \acrshort{rf} for early acceleration, \acrshort{srf} linacs and recirculating linacs, pulsed synchrotron lattices, rapid-ramping magnets, high-peak-power \acrshort{rf}, beam-loading control, and end-to-end longitudinal dynamics simulations.
Prototype hardware \acrshort{rnd} on a small pulsed dipole and a pulsed power source, to validate high field and high ramp rate material responses.
\acrshort{trl} assessments are shown in Table~\ref{T:Acceleration}.

\begin{table}[t]
\centering
\caption[Technical Readiness Levels -- rapid acceleration]{Technical Readiness Levels -- rapid acceleration.}
\label{tab:technology_assessment}
\begin{tabular}{p{64 mm}c p{79 mm}}
\toprule
\textbf{Key enabling \acrshort{rnd}} & \textbf{\acrshort{trl}} & \textbf{Comment} \\
\midrule
Pulsed magnet technology. & 4 &
Dominated by power converters, energy storage and the possible need for \acrshort{hts} dipoles. \\
\addlinespace
Evaluate fixed-field alternating-gradient accelerator options. & 2 &
Synergistic with generic community interest. \\
\addlinespace
End-to-end simulations. & 3 &
Preliminary parts of an integrated system already exist. \\
\addlinespace
High peak power \acrshort{rf}. & 3 &
New klystron or other tube designs may be needed for these unique requirements.  \\
\addlinespace
Rapid-ramping magnet demonstrator. & 4 & 
Straightforward but important. \\
\bottomrule
\end{tabular}
\label{T:Acceleration}
\end{table}

\paragraph{Discussion:}
Muons must be accelerated from a few MeV to multi-TeV energies within a timescale short compared to their relativistic lifetime, with the top-level parameter goals listed in Table~\ref{T:RMA}. 

\acrshort{srf} systems become viable when the beam energy reaches the few-hundred-MeV range. 
The reference design employs a staged approach -- an \acrshort{srf} linac followed by Recirculating Linear Accelerators that reach about 60~GeV. 
Longitudinal emittance is a critical parameter in reducing the decay of muons before higher-frequency \acrshort{rf} systems become effective. 
Pulsed synchrotrons become more efficient beyond about 60~GeV. 
The reference scenario employs a sequence of pulsed synchrotrons to accelerate to the highest energies.

A major technical risk lies in pulsed magnet technology, which must achieve rapid field ramps on millisecond timescales with high efficiency. 
U.S. efforts emphasize engineering-constrained designs, while international efforts (e.g., \acrshort{imcc}) focus on performance-driven concepts. 
Alternative concepts such as fixed-field alternating-gradient accelerators remain under study.

\subsubsection{Collider ring}

\paragraph{\acrshort{rnd} Priorities:}
Mature the design of collider ring optics and magnet systems for multi-TeV operation. 
Dynamic aperture and momentum-acceptance optimization, chromatic correction, beam-beam and impedance studies.
High-field, large-aperture \acrshort{sc} magnets with internal shielding or an open mid-plane design, cryogenic heat-load management, off-site neutrino radiation mitigation, and site-specific civil integration.
\acrshort{trl} assessments are shown in Table~\ref{T:Collider Ring}.

\begin{table}[htbp]
\centering
\caption[Technical Readiness Levels -- collider ring]{Technical Readiness Levels -- collider ring.}
\label{tab:technology_assessment}
\begin{tabular}{p{70 mm}c p{73 mm}}
\toprule
\textbf{Key enabling \acrshort{rnd}} & \textbf{\acrshort{trl}} & \textbf{Comment} \\
\midrule
Beam dynamics. & 4 &
Partially performed.  
Similar challenges with other collider designs. \\
\addlinespace
Large-aperture, high-field dipoles and quadrupoles. & 3 &
U.S. Magnet Development Program \cite{usmdp}. \\
\addlinespace
Niobium-tin magnets. & 6 &
Based on technology developed for High-Luminosity \acrshort{lhc}. 
Mostly value engineering. \\
\addlinespace
\acrshort{hts} magnets. & 3 &
Ongoing \acrshort{gard} effort. \\
\addlinespace
Beam-induced power deposition in magnets. & 3 &
Improved cooling and material studies needed. \\
\addlinespace
Off-site neutrino radiation. & 3 &
Potentially viable concepts developed. \\
\addlinespace
Demonstrator. & 3 &
Possible lower-energy implementation at first, producing physics results. \\
\bottomrule
\end{tabular}
\label{T:Collider Ring}
\end{table}

\paragraph{Discussion:}
Neutrino radiation and high-field magnet performance represent the most critical challenges. Energy acceptance and dynamic aperture limitations at the highest energies present significant but tractable risks. 
The absence of identified showstoppers, combined with steady progress, supports the conclusion that the reference collider ring concept is viable. 
The emphasis must shift toward engineering-level validation of key concepts, including magnet prototypes, shielding integration, and mover system implementation. 
Parameter goals for the collider ring are listed in Table~\ref{T:COLLIDER}.

A primary challenge results from the very short bunch length: a local chromaticity correction scheme using strong sextupoles near the Interaction Region (\acrshort{ir}) has been proposed, but higher-order chromaticities remain and limit momentum acceptance. 
High bunch intensity results in a large beam--beam parameter.
Transverse instabilities can be controlled with fast feedback systems. 
Dynamic aperture and energy acceptance require further optimization at the highest beam energies.

Collider ring magnets must combine high fields, large apertures, and substantial internal shielding while maintaining acceptable cryogenic loads. 
Beam-induced power deposition of approximately 500~W/m dominates the heat load in the cold mass. 
The reference design identifies a clear technology evolution from niobium-titanium and niobium-tin magnets towards \acrshort{hts} solutions (particularly \acrshort{rebco}) in higher-field applications.

The VIPER scheme for off- and on-site neutrino flux mitigation introduces vertical closed orbit excursions, adding complexity to lattice optics. 
Detailed \href{https://fluka.cern/}{FLUKA} simulations show that the VIPER mover system reduces neutrino flux in the arcs by approximately two orders of magnitude. 
Combined with optimized ring orientation and land-use planning, this approach brings off-site exposure levels well below regulatory goals. 
Preliminary site studies at \acrshort{cern} (using the \acrshort{sps} and \acrshort{lhc} tunnels) and \acrshort{fnal} (using the Tevatron tunnel) support the feasibility of this scheme, although detailed civil design work is still needed.

\subsubsection{Machine--Detector Interface}

\paragraph{\acrshort{rnd} Priorities:} 
Develop an integrated interaction region design that balances luminosity, detector performance, radiation protection, and engineering constraints. 
Final-focus optics with very small $\beta^*$, large-aperture radiation-resistant magnets, tungsten shielding liners, detector nozzle design, halo-loss control, vacuum and support systems.
Forward instrumentation, luminosity monitoring, and full accelerator-detector integration. 
A key challenge is the optimization of the nozzle geometry under engineering constraints set by radiation and heat load, requiring improvements to decay-background and halo simulations.  
\acrshort{trl} assessments are shown in Table~\ref{T:MDI}.

\begin{table}[htbp]
\centering
\caption[Technical Readiness Levels -- Machine--Detector Interface]{Technical Readiness Levels -- Machine--Detector Interface.}
\label{tab:technology_assessment}
\begin{tabular}{p{50 mm} c p{93 mm}}
\toprule
\textbf{Key enabling \acrshort{rnd}} & \textbf{\acrshort{trl}} & \textbf{Comment} \\
\midrule
Interaction region optics. & 4 &
Balance accelerator and detector needs and interests. \\
\addlinespace
Nozzle and vacuum systems. & 4 &
Radiation shielding challenge. \\
\addlinespace
Interaction region quads. & 3 &
\acrshort{gard} \acrshort{rnd}.  \acrshort{hts} and energy deposition challenges. \\
\addlinespace
Integration. & 3 &
Preliminary concepts exist. \\
\bottomrule
\end{tabular}
\label{T:MDI}
\end{table}

\paragraph{Discussion:}
The \acrshort{mdi} integrates the interaction region and detector designs to achieve a lattice with very small $\beta^*$ that keeps beam-induced backgrounds under control. 
The dominant background comes from muon decays, with additional contributions from pair production and halo losses.
The \acrshort{ir} needs tightly optimized masks and absorbers and a deep, nozzle-like shielding structure near the final-focus magnets, engineered for both radiation protection (including neutron moderation with borated material) and survivable heat loads. 
Meeting these goals also requires careful integration of detector nozzles and chambers and potential collimation and scraping systems, with strong coupling between lattice choices, magnet configuration, solenoid effects, and detector \acrshort{rnd} for background rejection.

The reference design features a compact interaction region with millimeter-scale $\beta^*$, surrounded by substantial shielding and a tapered detector nozzle. 
Thick tungsten liners within magnet apertures help to reduce heat loads and radiation damage, while upstream dipole chicanes intercept decay products to reduce background fluxes.

\clearpage
\subsection{Provide a technically informed, bounded estimate of the time required to complete each major \acrshort{rnd} category~\label{sec:detail_on_timeline}}

The Muon Collider is arguably the most technologically demanding accelerator facility ever conceived. 
Its realization will require advances across a broad spectrum of accelerator, magnet, target, \acrshort{rf}, and detector technologies, many of which must be demonstrated well beyond the current state of the art. 
Achieving the level of confidence needed for construction will require a coordinated program of progressively integrated technical demonstrations, each designed to reduce key technical risks and validate critical design assumptions.

The timeline in Table~\ref{T:TIMELINE} reflects the current \acrshort{trl}s.
It identifies opportunities to leverage ongoing \acrshort{rnd} programs, and estimates that a Technical Design Report could be delivered in approximately 20~years, assuming adequate resources and a technically limited \acrshort{rnd} plan.
Construction could begin after about 30~years.

\begin{table}[htbp]
\centering
\caption[Tentative technically limited timeline]
{Tentative technically limited timeline to complete major \acrshort{rnd} activities.
A Technical Design Report could be delivered in about 20~years, and construction could start in about 30~years.}
\begin{tabular}{c p{143 mm}}
\toprule
{\bf Phase}     & {\bf Activities} \\
{\bf [years]} \\

\midrule
{\bf 0--5} & {\bf Small-scale demonstrations} \\ 
\addlinespace 
& Test high-power target materials (solid, liquid metal, jets) \\
& Operate \acrshort{hts} magnets at high fields \\
& Test small-scale fast-ramping magnets for material and power source performance \\
& Operate \acrshort{rf} cavities in high magnetic fields, test cryogenic operation \\
& Integrate a component-level cooling cell (mechanical, cryogenics, couplers) \\
& Test space-charge mitigation and laser stripping for a proton driver \\
& Prototype beam halo scrapers and nozzle segments \\

\midrule 
{\bf 5--10} & {\bf Medium-scale demonstrations} \\ 
\addlinespace 
& Validate pion and muon yields with realistic targets \\
& Integrate a cooling-channel section (no beam $\rightarrow$ beam) \\
& Integrate a cooling cell \\
& Demonstrate efficient \acrshort{rf} sources and \acrshort{rf} power distribution \\
& Perform a medium-scale fast-ramping magnet string test \\
& Validate end-to-end simulations with hardware tests \\

\midrule
{\bf 10--20} & {\bf Large-system demonstrations} \\ 
\addlinespace 
& Operate a high-power target and capture solenoid \\
& Operate a partial cooling channel \\
& Prototype a rapid-cycling synchrotron or a fixed-field alternating-gradient accelerator section \\
& Demonstrate radiation, shielding, and backgrounds with an integrated beamline \\
& Operate storage ring dipole and final-focus magnets \\
\bottomrule
\end{tabular}
\label{T:TIMELINE}
\end{table}

\clearpage
\subsection{Assess the capabilities and roles of U.S. national laboratories in contributing to these critical areas~\label{sec:detail_on_capabilities}}

U.S. laboratories bring complementary expertise that spans most of the \acrshort{rnd} needed for a Muon Collider.
Four national laboratories -- \acrshort{bnl}, \acrshort{fnal}, \acrshort{lanl}, \acrshort{ornl} -- operate high-power proton beams where large-scale subsystem demonstrations could be performed. 
National laboratories also provide a vital ecosystem for attracting and training talent across research areas of broad applicability, including experiential opportunities for undergraduate students and graduate research appointments.

The potential roles of national laboratories are briefly summarized in Table~\ref{T:TON}.

\begin{table}[h]
\caption[Potential national laboratory roles]
{Summary of potential laboratory roles in Muon Collider accelerator \acrshort{rnd}. \label{tab:labrole}}
\label{T:TON}
\centering
\setlength{\tabcolsep}{4pt}
\begin{tabular}{lcccccccc}
\toprule
{\bf Topic}			& {\bf \acrshort{anl}}	& {\bf \acrshort{bnl}} & {\bf \acrshort{fnal}} & {\bf \acrshort{jlab}}   & {\bf \acrshort{lanl}}   & {\bf \acrshort{lbnl}} 	& {\bf \acrshort{ornl}}	& {\bf \acrshort{slac}}    \\
\midrule
Proton driver		& \ON 		& \ON 		& \ON 		& \ON          & \ON  		   & \ON 	      & \ON  		& -- \\
\acrshort{nc} \acrshort{rf} 				& \ON 		& \ON		& \ON 		& \ON          & \ON 		   & \ON          & \ON 		& \ON \\
\acrshort{srf} 				& \ON 		& -- 		& \ON 		& \ON          & -- 		       & \ON 	      & -- 			& -- \\
Magnets 			& \ON 		& \ON 		& \ON 		& -- 		   & -- 		       & \ON 		  & -- 			& -- \\
Cooling       		& -- 		& \ON		& \ON 		& \ON 		   & --             & \ON 	      & -- 			& \ON \\
Integration       	& -- 		& \ON		& -- 		& \ON			   & -- 		   & \ON 		      & -- 			& -- \\
Beam dynamics	    & \ON 		& \ON		& \ON 		& \ON 		   & \ON 		 & \ON            &\ON 			& \ON \\
Target, detector    & \ON 		& \ON 		& -- 		& -- 		   & \ON 		 & -- 		      & \ON 		& -- \\
Sims. \& models    	& \ON 		& \ON 		& \ON 		& \ON 		   & \ON 		 & \ON 		      & \ON 		& \ON \\
\bottomrule
\end{tabular}
\end{table}

At greater length, the national laboratory facilities and capabilities include:


\begin{description}[leftmargin=20 mm, style=nextline]

\item[\acrshort{anl}] Core areas of expertise include \acrshort{srf}, magnet research and development, targetry, normal-conducting (\acrshort{nc}) \acrshort{rf} high-gradient structures, high-power \acrshort{rf} generation, and high-performance computing. These capabilities are complemented by additional competency in materials testing, detector \acrshort{rnd}, low-frequency \acrshort{srf} cavities, and beam dynamics proton linacs. 

\item[\acrshort{bnl}] \acrshort{ags} for low-power front end and cooling channel mini-demonstrators.  Magnet \acrshort{rnd}, systems, and fabrication.  
Optics and beam dynamics, including beam dynamics experiments.  
Magnet test stands that may be augmented.
\acrshort{hts} and hybrid magnet \acrshort{rnd}, integration, detector, simulations and optical modeling.

\item[\acrshort{fnal}] \acrshort{srf} and magnet \acrshort{rnd} and fabrication. 
\acrshort{iota}, \acrshort{fast}, \acrshort{pipii} infrastructure. 
Magnet test stands that may be augmented.
Proton drivers, beamlines, cooling expertise, accelerator integration.
\acrshort{hpt} lab (material characterization equipment and advanced characterization tool development).
Advanced thermomechanical simulation expertise (including radiation damage, thermal shock and fatigue). Novel materials development and fabrication expertise.

\item[\acrshort{jlab}] \acrshort{srf} cavity and cryomodule design, fabrication and test capabilities, copper cavity fabrication capability, magnetron-based \acrshort{rf} systems.
Cavity and cryostat fabrication and cryogenic testing capabilities.
\acrshort{srf} cavity development, cooling,
\acrshort{rla} development, cryogenic materials testing.

\item[\acrshort{lanl}] Los Alamos Neutron Science Center (\acrshort{lansce}) proton accelerator for technology demonstration and radiation effects testing.
Target design, \acrshort{lace} injection, materials modeling, high-field normal-conducting pulsed magnets.

\item[\acrshort{lbnl}] Magnet \acrshort{rnd} facilities.  
Magnet design, \acrshort{hts} development.
Magnet test stands that may be augmented.
Facilities for designing and fabricating \acrshort{rf} cavities and modules for ionization cooling.
\acrshort{nc} \acrshort{rf} development.  \acrshort{srf} resonance control systems.  Proton front ends and laser stripping.  Beam dynamics expertise, modeling and high-performance software tools.  Start-to-end integrated simulations.

\item[\acrshort{nhmfl}]  HTS magnet R\&D, magnet test stands that may be augmented.

\item[\acrshort{ornl}] \acrshort{sns} facility for proton driver beam studies (e.g., bunch compression) and technology demonstrations (e.g., \acrshort{lace} injection). 
\acrshort{ornl} and \acrshort{sns} materials irradiation and damage testing capabilities.
\acrshort{sns} second target station for serial target tests.
Proton driver beam studies, materials irradiation, \acrshort{sns} target radiation testing, cooling channel materials studies, cooling channel prototype beamline.

\item[\acrshort{slac}] Facility to test \acrshort{rf} cavities in magnetic fields, \acrshort{rf} source test stands.
Facilities for sample and cavity testing including cryogenic operation, and a large-bore 5~T magnet undergoing commissioning.
Design, fabrication, and test capabilities for high peak-power \acrshort{rf} sources.
\acrshort{rf} testing in magnetic field, detector \acrshort{rnd}, \acrshort{rf} sources, \acrshort{rf} system integration, linac development.

\end{description}

\subsubsection{Near-term facility gaps and upgrade needs}

The near-term \acrshort{rnd} priorities take into consideration national lab capabilities, overlapping with laboratory programs and with generic benefit to accelerator development.
Upgrades needed to enable near- and medium-term \acrshort{rnd} include:
\begin{itemize}
    \item an \acrshort{rf} and magnetic field combined test stand, ideally with beam,
    \item an integrated cooling-channel test facility, and
    \item a high-power target test station with remote handling.
\end{itemize}

\clearpage
\section{Defining a demonstrator and test facility program}

A demonstrator and test facility program does not currently exist, but would be a key element of the \acrshort{rnd} strategy.
A long-term and clearly defined vision would guide technology development, establish priorities, and reduce technical risk.

Demonstrators and test facilities would address high-power proton drivers, target and capture systems, \acrshort{6d} muon cooling, rapid acceleration, high-field \acrshort{sc} magnets, and integrated machine operation. 
They would also provide experimental benchmarks for simulation and modeling tools.
A sustained demonstrator and test facility program would provide continuity for workforce development by creating a sequence of technically meaningful projects for students, early-career researchers, scientists, engineers, and technicians. The proposed IMCC ionization cooling demonstrator would be an element of this ensemble.

Reliance on a single large-scale demonstrator is unrealistic, especially in the absence of directed high-energy physics deliverables from that demonstrator. 
Rather, a coordinated sequence of tests would progressively validate the critical technologies and subsystems, and would provide a flexible and achievable path forward. 
Test results would deliver broadly relevant accelerator physics and technology outcomes and strengthen the case for continued investment.

A hierarchical model would progress from many small, targeted test facilities to a smaller number of increasingly integrated system-level demonstrations.
Early-stage, relatively low-cost tests and experiments would play a critical role, enabling rapid learning, providing data to benchmark simulations, and supporting informed down-selection among competing technical approaches. 
This reduces risk, accelerates progress, and avoids premature commitment to a single path before key uncertainties are resolved.

A distributed approach would create opportunities for broader national laboratory and university participation. 
For example, a targetry-focused test facility could be effectively established at \acrshort{bnl}, \acrshort{fnal}, \acrshort{lanl}, or \acrshort{ornl}. 
Multiple institutions could contribute meaningfully, reinforcing collaboration while advancing the overall \acrshort{rnd} program.

Theoretical and simulation work -- particularly in accelerator physics -- would be a critical component. 
Some recent studies provide useful direction, but many studies remain preliminary. 
Strengthening them while leveraging fast-paced progress in Artificial Intelligence (\acrshort{ai})-driven design platforms would help define realistic performance targets and would guide technology development priorities.

A well-defined roadmap would establish measurable technical milestones, align investments across national laboratories and universities, and provide clear decision points for future program optimization. 
An effective roadmap would align with broader accelerator and applied-technology programs, and would provide opportunities to leverage existing accelerator infrastructure. Demonstrators that also produce physics may be more appealing even if they require more than technology demonstrators alone.

A staged demonstrator and test facility program would provide a pathway from current \acrshort{rnd} activities to the integrated systems required for a future energy-frontier Muon Collider.

\clearpage
\section{Technologies}

The \acrshort{doe} \acrshort{gard} program is critical for advancing key technologies, such as high-field magnets, \acrshort{rf} systems, targets, beam dynamics, and novel acceleration techniques. \acrshort{gard} not only provides the technical basis for future large-scale facilities and improves the performance and cost-effectiveness of existing accelerators, but also fosters innovation with applications beyond physics, including medicine, industry, and other scientific fields.

\begin{table}[b!]
\centering
\caption[Superconducting magnet \acrshort{rnd} areas, challenges, and focus]
{Superconducting magnet \acrshort{rnd} areas and challenges.}
\begin{tabular}{lp{100 mm}}
\toprule
{\bf Target \& capture solenoids}   & 20~T, large-bore \\
~~Challenges:                         & High radiation, cooling, integration \\
~~\acrshort{rnd} focus:                         & \acrshort{hts}/low-temperature \acrshort{sc} (\acrshort{lts}) hybrid, thermal management, radiation resistance \\
{\bf Final cooling solenoid}        & $\geq$40~T, small bore \\
~~Challenges:                         & Extreme field, compactness \\
~~\acrshort{rnd} focus:                         & All-\acrshort{hts}/hybrid, quench protection, mechanical design \\
{\bf Fast-ramping magnets}             &0.5--2.0~T, 0.1--1~kHz \\
~~Challenges:                         & Rapid cycling, power efficiency  \\
~~\acrshort{rnd} focus:                         & Fast-ramping technologies, compatible power supplies \\
{\bf \acrshort{ir} quadrupoles}                    & 200--300~T/m, large aperture \\
~~Challenges:                         & Design concepts, luminosity impacts \\
~~\acrshort{rnd} focus:                         & Niobium-tin/\acrshort{hts} hybrid, thermal management, radiation resistance \\
{\bf Quench \& thermal management}         & \\
~~Challenges:                         & Detection, heat removal  \\
~~\acrshort{rnd} focus:                         & New quench detection methods, advanced cryogenics, thermal interface conductance  \\
{\bf Prototyping \& integration}        & \\
~~Challenges:                         & Large-scale \acrshort{rf} and magnet integration \\
~~\acrshort{rnd} focus:                         & Prototyping of integrated systems, manufacturability, system-level integration  \\
{\bf Radiation \& reliability}        & \\
~~Challenges:                         & High-radiation environment   \\
~~\acrshort{rnd} focus:                         & Materials development, shielding techniques, maintainability \\
\bottomrule
\end{tabular}
\label{T:MAGNETS}
\end{table}

\subsection{Superconducting magnets}

The magnet program benefits strongly from synergies with broader societal applications of \acrshort{hts} technology, including fusion energy and power systems. These synergies provide confidence that the required conductor and cable performance can be achieved within the timescales relevant to the collider program, provided adequate investment is maintained.
Magnet \acrshort{rnd} areas, parameters and challenges are listed in Table~\ref{T:MAGNETS}.

\subsection{\acrshort{rf} systems}

\acrshort{rf} systems span the entire machine from the proton driver through capture and cooling, post-cooling acceleration and storage. Leading parameters are listed in Table~\ref{T:RF}.

\begin{table}[b]
\centering
\caption[\acrshort{rf} systems]
{\acrshort{rf} systems parameters.}
\begin{tabular}{lccc}
\toprule
{\bf System}                & {\bf Frequency}   & {\bf Gradient}    & {\bf Power source} \\
                            &                   &                   & {\bf peak power} \\
                            & MHz               & MV/m              &    \\     
\midrule
{\bf Capture}               & 325--400           & $>$20            & Modest \\
{\bf Rectilinear cooling} & 325--650 & 30--50 & Very high   \\
{\bf Final cooling}      & 11--125 & 1--15 & Modest \\
{\bf Rapid acceleration}    \\
~~~Linac, normal-conducting & $<$650          & 30--50              & Very high\\
~~~Linac, superconducting   & $\sim$650          & 30--50              &Very high \\
~~~Rapid-cycling synchrotron, early, \acrshort{sc}     & $\sim$650         & 30--50         & Very high\\
~~~Rapid-cycling synchrotron, late, \acrshort{sc}      & $\sim$1,300       & 30--40         & High\\
\bottomrule
\end{tabular}
\label{T:RF}
\end{table}

Much progress was made through the U.S. Muon Accelerator Program (\acrshort{map}) \cite{Palmer:2013} and other studies in understanding the challenging requirements of these systems and developing designs to achieve them. 
Proof-of-concept tests of \acrshort{rf} in high magnetic fields demonstrated feasibility, but practical devices remain a long way off. Proton driver linacs exist that are close to the requirements for a Muon Collider and can benefit from material developments such as niobium-tin cavities for future upgrades. New and optimized designs are needed for all systems based on the concepts developed, engineered for integration into the overall complex.

\paragraph{Capture:} Low-frequency cavities are needed after the target to capture and phase rotate the beam. These must operate at high gradient and inside the solenoids that match into the cooling channel.

\paragraph{Rectilinear cooling:} The cooling \acrshort{rf} also operates in  strong magnetic fields that gradually increase along the channel. 
\acrshort{map} studies showed that \acrshort{rf} cavities with beryllium surfaces can withstand the high gradients and high magnetic fields up to at least 5~T, although beryllium is not ideal for large-scale installation. 
\acrshort{rnd} is needed to identify other suitable materials or alloys and other mitigations, such as cryogenic operation, that can meet these requirements. An aggressive \acrshort{rnd} plan would identify these solutions prior to finalizing the cooling cell design. A family of cooling cells is required with gradually reducing cell lengths as the beam emittance decreases. After the bunch merge the peak current is very high and the cavities are heavily beam loaded. 

\paragraph{Rapid acceleration:} Beam that survives the cooling channel must be accelerated as quickly as possible through a series of high-gradient cavities, starting with a low-frequency \acrshort{nc} linac. 
Acceleration switches to higher frequency \acrshort{sc} cavities placed between the focusing elements once a suitable energy is reached.
Later stages use an \acrshort{rla} to re-use the \acrshort{rf} sections. 
Finally, acceleration transitions to one or more rapid cycling synchrotrons (\acrshort{rcs}), doubling the frequency from about 650~MHz to about 1.3~GHz. 
These cavities are also heavily beam loaded. 
International Linear Collider (\acrshort{ilc})-level gradients are assumed for the 1.3 GHz cavities, although designs optimized for the \acrshort{rcs} need to be developed. New materials such as niobium-tin may give advantages.

\paragraph{Collider:} 
\acrshort{rf} systems in the collider need to handle high bunch charge, short lifetime, and frequent re-injection. 
Transient beam loading, impedances, and the high radiation environment may present significant challenges.

\paragraph {\acrshort{rf} power sources:} The capture and initial cooling systems require high peak power \acrshort{rf} sources to capture and accelerate the bunch train. 
After the merger only a single bunch is cooled and accelerated, and the peak \acrshort{rf} power increases. 
Very high peak power short pulse \acrshort{rf} sources need to be developed for a range of frequencies. The average power is modest due to the low repetition rate, although overall \acrshort{rf} efficiency is still important for limiting the peak demand from the modulators and power supplies. Recent developments in high-efficiency klystrons, multi-beam inductive output tubes, and magnetrons may be relevant.

\subsection{Targets}

The recent \acrshort{doe} \acrshort{hpt} \acrshort{rnd} roadmap~\cite{pellemoine2025high} identifies technology developments that are closely aligned with the needs of a future Muon Collider. 
The \acrshort{hpt} program addresses many of the critical challenges associated with operating multi-megawatt proton targets, including the development and qualification of novel target materials, characterization of irradiated materials, understanding thermal shock and fatigue, and prediction of material lifetime under extreme radiation environments. 
Complementary advances in multiphysics modeling are improving the ability to predict target performance, optimize designs, and reduce technical risk through validated simulations.

The \acrshort{hpt} roadmap also emphasizes the importance of advanced instrumentation for beam diagnostics and target health monitoring, alternative target concepts, post-irradiation examination, and specialized facilities to support target \acrshort{rnd}. 
These capabilities are directly applicable to the development of a Muon Collider target station and provide opportunities to leverage ongoing \acrshort{doe} investments. 
Close coordination between the Muon Collider program and the \acrshort{hpt} community will accelerate technology development, reduce duplication of effort, and strengthen the technical foundation for future high-power target systems.

\subsection{Beam simulations} 

Simulation plays a central cross-cutting role in every aspect of accelerator design. 
For example, beam transport, lattice design, magnet design, \acrshort{rf} design, and detector design all have their own specialized simulation needs.

Software readiness can be supported in part through continued (general) high-performance beam dynamics code development, support for the development of a community software ecosystem, and support for other collider modeling activities, including the Future Circular Collider (\acrshort{fcc}) and the Electron-Ion Collider (\acrshort{eic}). Important activities include:

\begin{itemize}
    \item extending existing simulation tools (e.g., code coupling for beam-matter interactions),
    \item developing a full end-to-end simulation framework (production $\rightarrow$ cooling $\rightarrow$ acceleration $\rightarrow$ detector),
    \item validating beam dynamics modeling results with experimental prototype data, and
    \item developing \acrshort{ai}-assisted agentic workflows for simulation and design.
\end{itemize}

The national laboratories host many of the original Muon Collider simulation experts.
Their expertise and knowledge need to be passed on to the next generation, who can 
collaborate with and learn from \acrshort{imcc} colleagues who did much of the recent work.
A U.S.-built Muon Collider requires ``in-house'' cultivation of this \acrshort{rnd} knowledge and expertise.

{\bf Beam dynamics} simulation studies are particularly crucial in knowledge transfer and future workforce cultivation.
It is essential to validate beam dynamics models against both existing facilities and proposed demonstrators.
Existing {\bf beam transport and lattice design} software tools include 
\href{https://blast.lbl.gov/codes/}{BLAST-WarpX/ImpactX}, \href{https://www.classe.cornell.edu/bmad/}{BMAD}~\cite{Sagan:Bmad2006}, \href{http://g4beamline.muonsinc.com/}{G4BeamLine}~\cite{g4beamline}, 
\href{https://madx.web.cern.ch/}{MAD-X}~\cite{MADX:2024}, 
\href{https://pyorbit.readthedocs.io/}{PyORBIT}~\cite{shishlo2015particle}, 
\href{https://xsuite.readthedocs.io/en/latest/#}{XSuite}~\cite{Iadarola:2023fuk}, ACE3P~\cite{candelhigh,ace3p}, and specialized codes such as 
\href{https://bdsim-collaboration.github.io/web/}{BDSIM}~\cite{Boogert:2013},
\href{https://gitlab.cern.ch/rf-track}{RF-Track}~\cite{latina:hb2023-wea3c1}, and
\href{https://abpcomputing.web.cern.ch/codes/codes_pages/Guinea-Pig/}{GUINEA-PIG}~\cite{Schulte:1999}. 
Muon Collider beam transport is unusual compared to typical collider beam transport due to the frequent beam-matter interactions that play a central role in ionization cooling.  As a result, cooling channel simulations require coupling to beam-matter interaction models such as those in \href{https://fluka.cern/}{FLUKA} and \href{https://geant4.web.cern.ch/}{GEANT4}.

Specialized software tools must be used in several \acrshort{rnd} areas. 
For example, the close proximity of conducting structures (\acrshort{rf} and foils) in the cooling channels motivates the use of fully electromagnetic {\bf particle-in-cell} simulations to understand the effects of beam loading and conducting boundary conditions.  
{\bf Wakefields and collective effects} need to be modeled in the cooling channel and acceleration chain to understand the interplay between beam--beam and wakefield effects in the collider ring. 

These \acrshort{rnd} areas are connected to a broader accelerator community need for integrated software workflows that facilitate multiphysics.
{\bf End-to-end modeling} involves coupling multiple codes and requires the efficient use of high-performance computing and graphics processing unit resources.  
There is also an increasing drive toward the use of {\bf differentiable simulation} tools that facilitate gradient-based optimization and sensitivity studies, fast {\bf neural-network surrogate} models, and {\bf \acrshort{ai}-assisted} workflows.  

\clearpage
\section{Surviving the Valley of Death: Synergies and Alternative Support for Technology Development} 

A combination of coordinated federal \acrshort{rnd} programs, laboratory capabilities, and university partnerships could create a sustainable innovation ecosystem supporting progress toward a Muon Collider.  This would simultaneously advance accelerator science, develop a highly skilled workforce, and deliver broad benefits across many applications of advanced accelerator technologies.

\begin{figure}[htbp]
    \centering
    \includegraphics[width=0.85\textwidth]{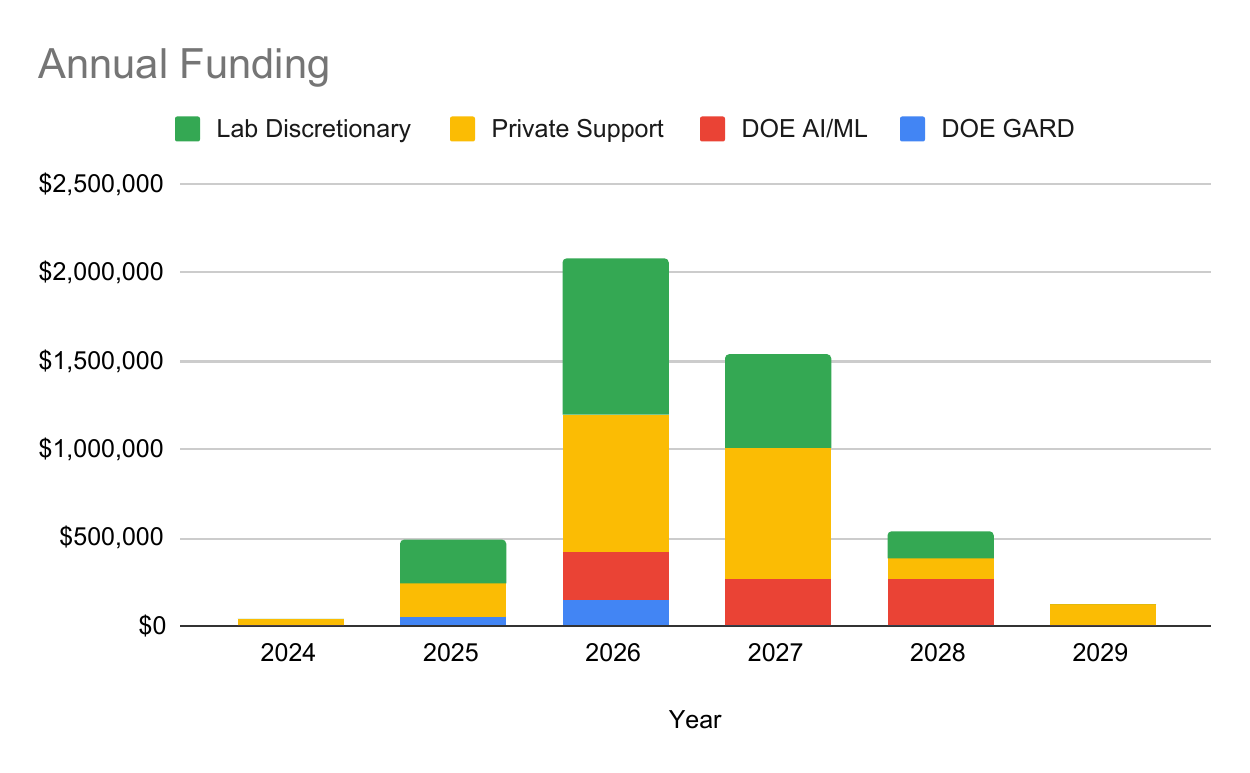}
    \caption[Recent funding trends in Muon Collider technology development]
    {Recent funding trends in Muon Collider technology development, including targeted accelerator \acrshort{rnd}, Artificial Intelligence/Machine Learning (\acrshort{aiml}) initiatives, university research grants, and \acrshort{ldrd} activities. (Courtesy USMCC)}
    \label{fig:muon_funding}
\end{figure}

However, Figure~\ref{fig:muon_funding} shows the very limited funding in recent years, and suggests that the near-term future will be even more challenging. 
Below we clarify some ongoing and potential near-term opportunities that could help grow the U.S. \acrshort{rnd} efforts. These could perhaps be funded by \acrshort{gard}, but any enhanced investment across these areas, such as sustained Laboratory Directed Research and Development (\acrshort{ldrd}), would be important in maintaining progress toward a Muon Collider.

\subsection{\acrshort{rf} systems}

\acrshort{rf} system requirements for high-gradient, high-efficiency acceleration align closely with developments for other future colliders (e.g., \acrshort{fcc}, \acrshort{eic}), high-intensity proton facilities, advanced light sources, and energy-recovery linacs. Innovations in \acrshort{srf}, \acrshort{nc} \acrshort{rf}, high-power \acrshort{rf} sources, and cryomodule efficiency can be jointly pursued, benefiting many accelerator science communities.
These shared needs naturally connect to efforts supported by the \acrshort{doe} \acrshort{gard} program, which fosters cross-cutting advances in \acrshort{rf} technology and enables collaboration across accelerator applications.

\subsection{Superconducting magnets and fusion technology}

Magnet development is also a major area of synergy. Muon Colliders require high-field, rapidly ramped, and radiation-tolerant magnets, overlapping significantly with the needs of fusion energy systems, which demand robust, high-field \acrshort{sc} magnets with reliable and repeatable performance.

Progress in \acrshort{hts} conductor technology, quench protection, and manufacturing methods can therefore be leveraged across broad domains, including applications such as medical and industrial accelerators. 
In addition, muon production targets inherently generate intense secondary particle fields, enabling concepts for volumetric neutron sources that are directly relevant to fusion materials testing.

There is a natural synergy between Muon Collider target systems and the development of irradiation environments needed for fusion energy \acrshort{rnd}.

\subsection{Beam dynamics, controls, and Artificial Intelligence}

Beam dynamics, controls, and end-to-end system optimization benefit from emerging \acrshort{aiml} techniques. 
The complexity of Muon Collider systems -- spanning particle production, cooling, acceleration, and collision -- drives the need for integrated, high-fidelity simulations and real-time optimization.

The use of \acrshort{aiml} tools and workflows assisted by agentic large language models plays an increasingly critical role in accelerator science and technology, including:  
\begin{itemize}
\item beam dynamics simulation and software development,
\item accelerator control using digital twins, and
\item accelerator design.  
\end{itemize}
The role of \acrshort{aiml} will increase significantly as part of the U.S. GENESIS Mission, which includes ``Enhancing Particle Accelerators for Discovery" as one of its 26 National Science and Technology Challenges~\cite{GENESIS}.  
In preparation for this, the Muon Collider community has recently engaged with specialists in \acrshort{ai} methods for accelerator and detector applications as part of a two-day workshop~\cite{AI4MuC}.

These same \acrshort{ai}-enabled approaches are increasingly important for light sources, fusion devices, and other large-scale scientific facilities. 
Investments in \acrshort{ai} for surrogate modeling, digital twins, and autonomous operation will deliver broad benefits across multiple mission areas.
Universities will play a key role in advancing these capabilities through fundamental research programs, algorithm development, and the training of the next generation of accelerator scientists and engineers.

\clearpage
\subsection{Currently or previously operating muon facilities}

\begin{table}[b!]
\centering
\caption[Current and previous muon facilities]
{A partial list of current and previous muon accelerators and experiments.
}
\begin{tabular}{lcccp{0.42\textwidth}}
\toprule
{\bf Accelerator}  & {\bf Type}    & {\bf Proton}  & {\bf Rep.} & {\bf Comment} \\
{\bf or laboratory} &              & {\bf energy}  & {\bf rate}   \\
                &           & [GeV] &  [Hz] \\
\midrule
{\bf Current} \\
\acrshort{csns}, China & Synchrotron & 1.6 & 25 & \acrshort{melody} beamline, \acrshort{msr} etc.
\\
\acrshort{fnal}, U.S. & Synchrotron & 8.0 & 15 & g-2 \& (soon) $\mu2e$ experiments \\
\acrshort{isis}, UK & Synchrotron   & 0.8      & 50 & MuX, CHRONUS \& ARGUS beamlines\\
\acrshort{jparc}, Japan & Synchrotron & 3.0   & 25 & \acrshort{msr} beamline, COMET experiment \\
\acrshort{psi}, Switzerland & Cyclotron & 0.59 & \acrshort{dc} & $\pi$E1, $\pi$E3, $\mu$E1 \& (soon) $\mu$3e beamlines \\
\acrshort{triumf}, Can. & Cyclotron & 0.52   & \acrshort{dc} & M20, M9, M15, \& M11 beamlines \\
{\bf Previous} \\
\acrshort{bnl}, U.S. & Synchrotron & 24 & 0.4 & g-2 before removal to \acrshort{fnal} \\
\acrshort{cern}, Switz. & Synchrotron & 26 & 0.4 & Storage ring measurements, cf. g-2 \\
\acrshort{kek}, Japan & Synchrotron & 0.5 & 20 & Superseded by \acrshort{jparc} MUSE \\
\acrshort{lampf}, U.S. & Linac & 0.8 & 120 & High-intensity \acrshort{msr}, muon decay \\
\bottomrule
\end{tabular}
\label{T:OTHER}
\end{table}

Several accelerators currently generate -- or previously generated -- muons for materials science, basic research, or particle beam experiments.
Table~\ref{T:OTHER} lists the major facilities and experiments that are (or were) built around muon beams -- both secondary muon beams produced as a byproduct of proton accelerators and dedicated
muon-storage-ring efforts. 
The list is not exhaustive.
It does not include every small university-scale beamline or historical bubble-chamber muon exposure.

Muon Spin Rotation, Relaxation, and Resonance (\acrshort{msr}) experiments continue to be active, for example, in what is generally considered the world-leading \acrshort{dc} muon source at \acrshort{psi}.
\acrshort{isis} operates one of the highest-intensity pulsed muon beams.
TRIUMF's continuous-beam muon source was historically central to \acrshort{msr} development.

\subsection{Muon-catalyzed fusion facility}

An attractive opportunity within a staged U.S. demonstrator program is the development of a muon-catalyzed fusion (\acrshort{mcf}) research facility. 
Although the scientific objectives of a \acrshort{mcf} facility differ from those of a Muon Collider, many of the enabling technologies are directly relevant. 

A \acrshort{mcf} facility would require the production, capture, transport, and manipulation of low-energy muon beams, together with proton drivers, high (peak) power targets, advanced beam diagnostics, precision timing, cryogenic systems, and sophisticated controls. 
Such a facility would provide an integrated environment for developing and validating many of the subsystems that are essential for a Muon Collider, while operating at significantly lower average powers, with correspondingly lower costs and technical risks.

A \acrshort{mcf} facility would also provide an important platform for systems integration, operational experience, and workforce development. 
It would enable researchers to gain experience with high-intensity muon beam production, target operation, radiation shielding, remote handling, detector systems, and machine controls in a realistic operational environment. 
These capabilities would help retire technical risks associated with future Muon Collider demonstrators while providing valuable scientific opportunities in muon-induced processes and fusion science. 

A \acrshort{mcf} facility could therefore serve as an intermediate demonstrator in a staged Muon Collider \acrshort{rnd} roadmap, simultaneously advancing fundamental science and accelerator technology while developing the needed integrated engineering expertise.
When combined with other staged demonstrators targeting ionization cooling, rapid acceleration, and collider subsystems, a \acrshort{mcf} facility would provide a complementary pathway for validating critical technologies while broadening the scientific impact and expanding the muon-accelerator community.

Other muon accelerator concepts that have significant synergies with Muon Collider \acrshort{rnd} include \acrshort{mic}~\cite{Acosta2021} and \acrshort{mutristan}~\cite{Hamada2022}.

\clearpage
\section{Coordination Across Countries, Labs, and Universities}

\subsection{International coordination with the \acrshort{imcc}}

The development of a Muon Collider is inherently an international endeavor. No single institution or nation possesses all of the expertise, facilities, and resources required to address the full range of scientific and technical challenges. The \acrshort{imcc}~\cite{imcc}, coordinated through \acrshort{cern}, provides a framework for integrating worldwide expertise and developing a common technical basis for evaluating the feasibility of a future Muon Collider.

Through coordinated working groups spanning the full accelerator complex -- from the proton driver, target, capture, and cooling systems to acceleration, the collider ring, detectors, and the machine--detector interface -- the \acrshort{imcc} enables the international community to share technical knowledge, coordinate design activities, avoid unnecessary duplication, and converge on common assumptions, requirements, and performance goals.

Within this framework, the United States can provide major contributions through the capabilities of its national laboratories and universities. Conversely, there could be international contributions to US-hosted demonstration efforts. U.S. institutions possess leading expertise in accelerator physics, \acrshort{sc} magnets, high-gradient \acrshort{rf} systems, \acrshort{hpt}, beam dynamics, muon cooling, advanced simulation, and large-scale systems engineering. A clearly defined U.S. demonstrator and test facility roadmap would complement the broader \acrshort{imcc} program by providing opportunities to develop and validate critical technologies that can be incorporated into the international design effort.

Close coordination among U.S. national laboratories, \acrshort{cern}, European, and Asian partner laboratories within the \acrshort{imcc} would maximize the scientific return on investments and ensure that national activities contribute effectively to a common international strategy. 
Coordination would naturally include the development of shared technical requirements, compatible interfaces and standards, joint simulation and design efforts, coordinated use of specialized test facilities, personnel exchanges, and, where appropriate, jointly developed demonstrators and test facilities.

Continued engagement with Europe, Asia, and the \acrshort{imcc} would also help maintain programmatic momentum, broaden the technical workforce, and establish the collaborative relationships needed for a future project. 
A strong and clearly articulated U.S. program, aligned with the \acrshort{imcc}, would retain responsibility for identifiable technical deliverables, positioning the U.S. as a leading partner in Muon Collider development.

\subsection{Coordination across U.S. universities}

Previous and ongoing U.S. initiatives -- the Muon Accelerator Program (2010--2017)~\cite{Palmer:2013} and the U.S. Muon Collider Collaboration (\acrshort{usmcc}) -- have already been important catalysts for workforce development. 
They invigorated university engagement in accelerator science, attracting faculty, graduate students, and postdoctoral researchers from both accelerator physics and the broader experimental and theoretical particle physics communities. 
This pipeline of academic participation fostered closer collaborations between U.S. universities and national laboratories.

The \acrshort{usmcc} further strengthened these partnerships through programs such as µCATALYST, which brings university students to national laboratories each summer to work alongside laboratory scientists and engineers on projects directly supporting Muon Collider \acrshort{rnd}.
The increased engagement of students and early-career researchers was reflected in growing participation at Muon Collider workshops, collaboration meetings, and accelerator conferences, helping to ensure the long-term vitality of the U.S. accelerator workforce.

University groups now contribute to essentially every aspect of the Muon Collider and demonstrator programs, including the proton driver, target and capture systems, ionization cooling, acceleration, beam dynamics, \acrshort{sc} magnets, collider ring, \acrshort{mdi}, and the simulation and modeling tools that support system optimization. 
Their activities combine the complementary strengths of national laboratories with the research environment, innovation, and educational mission of universities. 

Universities also provide resources that complement \acrshort{doe} investments.
Many university groups leverage institutional support, interdisciplinary research initiatives, and private foundation funding to advance key technology developments. 
This broad support expands the technical expertise available to Muon Collider \acrshort{rnd}, while also providing outstanding training opportunities for the next generation of accelerator scientists and engineers.

\clearpage
\section{Acknowledgments}

The authors thank the members of the International Muon Collider Collaboration for valuable discussions. 
In particular, we thank Kavin Ammigan, Sergo Jindariani, Daniel Schulte, Steinar Stapnes, and Katsuya Yonehara.
Much of the input for this report was derived from a series of technical seminars listed in Table~\ref{T:seminar}. The study group extends its gratitude to the speakers and community participants for their valuable contributions.

Finally, the following individuals have provided valuable input on the draft report -- 
\acrshort{anl}: Brahim Mustapha, John Power; 
\acrshort{bnl}: Scott Berg, Stephen Brooks, Kevin Brown, Dmitri Denisov, Kiel Hock, Haixin Huang, Mithlesh Kumar, Brendan Lepore, Michiko Minty, Vikas Teotia;
\acrshort{cern}/\acrshort{imcc}: Daniel Schulte, Steinar Stapnes; \acrshort{fnal}: Norbert Holtkamp, Sergo Jindariani;
\acrshort{lbnl}: Cameron Geddes;
\acrshort{stfc}: Chris Rogers;
\acrshort{utk}: Tova Holmes.

\begin{table}[ht]
\centering
\caption[Seminar schedule]
{Seminar speakers and topics in 2025 and 2026.}
\begin{tabular}{ccl l}
\toprule
{\bf \#}                & {\bf Date}   & {\bf Speaker}    & {\bf Topic} \\
\midrule
{\bf 1}               & 27-Oct-2025           & M. Palmer            & Overview of technological challenges \\
{\bf 2} & 03-Nov-2025 & K. Yonehara & Target station   \\
{\bf 3} & 10-Nov-2025 & J. Eldred& Challenges for the proton complex\\
{\bf 4} & 17-Nov-2025 & D. Schulte \& C. Ahdida & Collider   \\
{\bf 5} & 24-Nov-2025 & A. Lechner & Machine--Detector Interface   \\
{\bf 6} & 01-Dec-2025 & S. Berg & Acceleration   \\
{\bf 7} & 08-Dec-2025 & C. Rogers & \acrshort{6d} cooling\\
{\bf 8} & 05-Jan-2026 & R. Taylor & Final cooling   \\
{\bf 9} & 12-Jan-2026 & T. Luo & Normal-conducting \acrshort{rf}   \\
{\bf 10} & 26-Jan-2026 & L. Bottura & Magnet system for muon capture and target   \\
{\bf 11} & 02-Feb-2026 & S. Belomestnykh & Superconducting \acrshort{rf}   \\
{\bf 12} & 16-Feb-2026 & D. Stratakis & Demonstrator and \acrshort{rf} test stands   \\
{\bf 13} & 02-Mar-2026 & K. Long & Lessons learned from \acrshort{mice} \\
{\bf 14} & 09-Mar-2026 & S. Mariotto \& B. Caiffi & Collider ring magnets   \\
{\bf 15} & 16-Mar-2026 & F. Boattini & Rapid Cycling Synchrotrons   \\
{\bf 16} & 23-Mar-2026 & K. Badgley & Solenoids   \\
{\bf 17} & 06-Apr-2026 & S. Pagan Griso & Detector challenges   \\
\bottomrule
\end{tabular}
\label{T:seminar}
\end{table}

Generative artificial intelligence was used as an editorial assistant during the preparation of this report to improve readability, organization, and clarity. 
The authors developed, reviewed, and verified all content, and assume full responsibility for the accuracy and integrity of the document.

This work was supported by the U.S. Department of Energy, Office of Science or National Nuclear Security Administration, under contract numbers  
DE-AC02-06CH11357 (\acrshort{anl}), 
DE-SC0012704 (\acrshort{bnl}),
89243024CSC000002 (\acrshort{fnal}), 
89243126CSC000213 (\acrshort{jlab}), 
89233218CNA000001 (\acrshort{lanl}),
DE-AC05-00OR22725 (\acrshort{ornl}), and
DE-AC02-76SF00515  (\acrshort{slac}).

\clearpage
\section*{Appendices}
\addcontentsline{toc}{section}{Appendices}

\appendix
\section{Existing International Efforts}

\subsection{International Muon Collider Collaboration}

The \acrshort{imcc}~\cite{imcc} was established in 2021 under \acrshort{cern} coordination to provide a unified international framework for assessing the feasibility of a Muon Collider. Building on the U.S. Muon Accelerator Program (2010--2017)~\cite{Palmer:2013}, the \acrshort{imcc} consolidates global expertise, harmonizes design assumptions, and evaluates technical challenges in a coordinated manner. It is explicitly exploratory, organized into subsystem-focused working groups covering the proton driver, target systems, muon cooling, acceleration, collider ring, \acrshort{mdi}, and detectors. To date, progressively more integrated design studies have not identified fundamental physical or technological barriers to feasibility.

\subsection{MuCol}

MuCol is a comprehensive design study project funded by the European Union under Horizon Europe to develop a high-energy Muon Collider design and evaluate its feasibility.
It serves as the main contribution to ongoing global Muon Collider \acrshort{rnd} efforts outside of \acrshort{cern}.
It addresses muon production and capture, ionization cooling, rapid acceleration systems, high-field and fast-ramping magnets, \acrshort{mdi}, detector backgrounds, and overall facility integration, as well as practical considerations such as power consumption, environmental impact, and cost optimization.

\subsection{U.S. Muon Collider Collaboration}

The U.S. Muon Collider Collaboration builds on a long history of muon collider studies dating back to the 1990s.  \acrshort{map} (2010--2017), jointly led by \acrshort{fnal} and \acrshort{lbnl}, developed much of the conceptual framework underlying modern Muon Collider designs and provided critical experimental validation through \acrshort{mice} and the MERIT liquid-metal target experiment at \acrshort{cern}~\cite{McDonald:2009zz}. Although \acrshort{map} concluded in 2017, it left behind a mature conceptual foundation and a trained technical community. 

Today the collaboration brings together nearly 300 members from universities and national laboratories, spanning accelerator science, experiment, and theory, with approximately one-third of its membership consisting of early-career researchers. Its core objectives are to define the R\&D program needed to inform the mid-P5 assessment; lead the design of a U.S.-based muon cooling demonstrator; engage and coordinate with the international community; develop a long-term U.S. vision toward a future muon collider; and strengthen the theory-driven physics case. 

The USMCC is pursuing a broad program aimed at addressing key technical and scientific questions on the path toward a muon collider. Accelerator activities include development of end-to-end design and simulation capabilities; studies of high-intensity proton accumulation and compression; R\&D on high-power targets, magnets, and high-gradient RF systems; and development of a conceptual design for a muon cooling technology demonstrator. The U.S. community is also making substantial contributions to detector and computing studies, including tracking, calorimetry, detector technologies, real-time data acquisition, and AI/ML, while the theory community continues to develop the collider physics case and explore additional opportunities enabled by intense proton and muon beams. Workforce development is an important component of the program, including the µCATALYST initiative, which supports graduate students working on muon collider accelerator R\&D across U.S. national laboratories. 

The USMCC is intended to complement and strengthen the international Muon Collider program rather than duplicate it, organizing U.S. expertise and resources around areas where the United States can make high-impact contributions. In the near term, the emphasis is on establishing the R\&D priorities, demonstrating critical technologies, developing the scientific case, and building the community and expertise needed to inform future decisions about the path toward a muon collider.

\subsection{Current status and outlook}

The \acrshort{imcc}, MuCol, and \acrshort{usmcc} collectively represent a focused feasibility and risk-reduction effort. Integrated design studies indicate that the remaining challenges are predominantly engineering and system-integration issues rather than fundamental limitations of physics or technology. These activities provide a credible and steadily advancing technical foundation for evaluating the Muon Collider as a potential future energy-frontier facility, and support informed decision-making regarding the potential evolution from feasibility studies toward a decision-ready concept.

\clearpage
\section{Charge to the National Lab Accelerator Study Group for a Muon Collider}
The National Lab Accelerator Study Group for a Muon Collider was formally established in August 2025 at the request of Dr. Young-Kee Kim, Professor at the University of Chicago, who at the time was the interim Director of Fermi National Accelerator Laboratory. The study group brings together eight national laboratories that operate accelerators.  

It was tasked with answering the following charge questions approved by the participating laboratory directors:
\begin{enumerate}
    \item Identify the key accelerator challenges that must be addressed to enable the realization of a 10\,TeV Muon Collider;
    \item Evaluate the \acrshort{rnd} steps, including demonstrations, simulation and software tools, and facility capabilities required to address these challenges;
    \item Provide a technically informed, bounded estimate of the time required to complete each major \acrshort{rnd} category, recognizing that these estimates will inform -- but not constitute -- a full collider development plan; and
    \item Assess the capabilities and roles of U.S. national laboratories in contributing to these critical areas.
\end{enumerate}

{\it ``The study should build on existing U.S. and International Muon Collider Collaboration (\acrshort{imcc}) \acrshort{rnd} planning documents, critically evaluating them to identify areas of alignment, overlap, and gaps that require further attention. 
This includes technical gaps, infrastructure needs, and coordination challenges.

The group is not tasked with developing a full collider concept or roadmap, but rather with clarifying and prioritizing the foundational technical work needed over the next decade. 
The resulting estimates and findings should serve as inputs for a coordinated, evolving timeline for muon collider \acrshort{rnd}.

Strong emphasis should be placed on coordination across U.S. national labs, as well as alignment with international efforts, to ensure that the U.S. program is both complementary and high impact in the global context.''}

\paragraph{Membership.}
The study group includes representatives from each of eight national laboratories: Philippe Piot from Argonne National Laboratory, Chad Mitchell from Lawrence Berkeley National Laboratory, Stephen Peggs from Brookhaven National Laboratory, Steve Gourlay from Fermi National Accelerator Laboratory (Chair), Robert Rimmer from Thomas Jefferson National Accelerator Facility, Steve Russell from Los Alamos National Laboratory, Fulvia Pilat from Oak Ridge National Laboratory, and Emilio Nanni from \acrshort{slac} National Accelerator Laboratory.
\clearpage

\printglossary[title=List of Acronyms, toctitle=List of Acronyms, type=\acronymtype]

\clearpage
\bibliographystyle{unsrturl}     
\addcontentsline{toc}{section}{References}
\bibliography{references}     

@misc{imcc,
    url = {https://muoncollider.web.cern.ch/},
    title = {{International Muon Collider Collaboration (IMCC)}}
}

@book{BejarAlonso:2749422,
    address = {Geneva},
    author = {B{\'e}jar Alonso, I. and Bottura, L and Apollinari, G and Br{\"u}ning, O and Rossi, L and Zimmermann, F},
    doi = {10.23731/CYRM-2020-0010},
    publisher = {CERN},
    series = {CERN Yellow Reports: Monographs},
    title = {{High-Luminosity Large Hadron Collider (HL-LHC): Technical Design Report}},
    year = {2020},
    url = {https://doi.org/10.23731/CYRM-2020-0010}
}

@misc{acciarri2016_short,
    archiveprefix = {arXiv},
    author = {Acciarri, R. and others},
    eprint = {1601.05471},
    primaryclass = {physics.ins-det},
    title = {{Long-Baseline Neutrino Facility (LBNF) and Deep Underground Neutrino Experiment (DUNE) Conceptual Design Report Volume 1: The LBNF and DUNE Projects}},
    url = {https://arxiv.org/abs/1601.05471},
    year = {2016}
}

@misc{consolidatedParameters,
    author = {Taylor, R. and Chance, Antoine and Giove, Dario Augusto and Milas, Natalia and Losito, Roberto and Lucchesi, Donatella and Rogers, Chris and Rossi, Lucio and Schulte, Daniel},
    doi = {10.5281/ZENODO.17476875},
    publisher = {Zenodo},
    title = {Consolidated Parameters},
    url = {https://zenodo.org/doi/10.5281/zenodo.17476875},
    year = {2025}
}

@article{mice_cooling_2020,
    author = {M. Bogomilov and others (MICE Collaboration)},
    doi = {10.1038/s41586-020-1958-9},
    journal = {Nature},
    pages = {53--59},
    title = {First demonstration of ionization cooling by the Muon Ionization Cooling Experiment},
    url = {https://www.nature.com/articles/s41586-020-1958-9},
    volume = {578},
    year = {2020}
}

@article{Belosevic2019muCool,
    author = {Belosevic, Ivana and others},
    doi = {10.1140/epjc/s10052-019-6932-z},
    journal = {The European Physical Journal C},
    number = {5},
    pages = {430},
    title = {mu{C}ool: a next step towards efficient muon beam compression},
    url = {https://doi.org/10.1140/epjc/s10052-019-6932-z},
    volume = {79},
    year = {2019}
}

@inproceedings{Curco:2024,
    author = {Curc{\`o}, C. and others},
    booktitle = {{14th International Particle Accelerator Conference}},
    title = {{FLUKA simulations of neutrino-induced effective dose at a Muon Collider}},
    url = {https://cds.cern.ch/record/2912496},
    year = {2024}
}

@inproceedings{Palmer:2013,
    author = {Palmer, M. A. and others},
    booktitle = {{2013 International Workshop on Beam Cooling and Related Topics}},
    title = {{An Overview of the US Muon Accelerator Program}},
    url = {https://proceedings.jacow.org/COOL2013/papers/moam2ha02.pdf},
    year = {2013}
}

@inproceedings{Boogert:2013,
    author = {Boogert, S. T. and others},
    title = {{Beam Delivery Simulation (BDSIM): A GEANT4 Based Toolkit for Diagnostic and Loss Simulation}},
    booktitle = {2013 International Beam Instrumentation Conference},
    year = {2013},
    url = {https://epaper.kek.jp/IBIC2013/papers/wepc46.pdf}
}

@misc{g4Beamline,
    author = {{Muons Inc.}},
    howpublished = {\url{http://g4beamline.muonsinc.com/}},
    title = {{G4Beamline} Particle Tracking Simulation Program}
}

@article{Sagan:Bmad2006,
    author = {Sagan, D.},
    doi = {10.1016/j.nima.2005.11.001},
    journal = {Nucl. Instrum. Meth.},
    number = {1},
    pages = {356-359},
    title = {{BMAD: A relativistic charged particle simulation library}},
    volume = {A558},
    year = {2006}
}

@article{Iadarola:2023fuk,
    archiveprefix = {arXiv},
    author = {Iadarola, Giovanni and others},
    doi = {10.18429/JACoW-HB2023-TUA2I1},
    eprint = {2310.00317},
    journal = {JACoW},
    pages = {TUA2I1},
    primaryclass = {physics.acc-ph},
    title = {{XSuite: An Integrated Beam Physics Simulation Framework}},
    volume = {HB2023},
    year = {2024}
}

@inproceedings{MADX:2024,
    author = {Fuster Martinez, N.},
    booktitle = {Proceedings of the Joint Universities Accelerator School (JUAS): Courses and tutorials},
    doi = {10.23730/CYRSP-2024-003.257},
    editor = {M{\'e}tral, E.},
    note = {CERN Yellow Reports: School Proceedings},
    title = {{MAD-X} simulation code},
    url = {https://cds.cern.ch/record/2928179},
    year = {2024}
}

@techreport{shishlo2015particle,
    author = {Shishlo, Andrei and Cousineau, Sarah and Holmes, Jeffrey and Gorlov, Timofey},
    institution = {Oak Ridge National Laboratory (ORNL), Oak Ridge, TN (United States)},
    month = {January},
    title = {{The Particle Accelerator Simulation Code PyORBIT}},
    url = {https://www.osti.gov/biblio/1185836},
    year = {2015}
}

@misc{GENESIS,
    author = {{U.S. Department of Energy}},
    title = {{Genesis Mission National Science and Technology Challenges}},
    url = {https://www.energy.gov/documents/genesis-mission-science-and-technology-challenges},
    year = {2026}
}

@misc{AI4MuC,
    title = {{Artificial Intelligence for Muon Colliders (AI4MuC) Virtual Mini-Workshop}},
    url = {https://indico.muoncollider.us/event/41/},
    year = {2026}
}

@article{Hwang_2001,
    doi = {10.1143/JJAP.40.7140},
    url = {https://doi.org/10.1143/JJAP.40.7140},
    year = {2001},
    month = {December},
    volume = {40},
    number = {12R},
    pages = {7140},
    author = {Hwang, Il-Moon and Kim, Eun-San and Yoon, Moohyun},
    title = {Strong--Strong Beam--Beam Simulation for a Muon-Collider Ring},
    journal = {Japanese Journal of Applied Physics}
}

@article{PhysRevAccelBeams.26.043501,
    title = {{Experimental demonstration of sequential excitation scheme for ${\mathrm{H}}^{\ensuremath{-}}$ laser assisted charge exchange}},
    author = {Aleksandrov, Alexander and Cousineau, Sarah and Gorlov, Timofey and Liu, Yun and Oguz, Abdurahim and Shishlo, Andrei and Zhukov, Alexander and Kay, Martin},
    journal = {Phys. Rev. Accel. Beams},
    volume = {26},
    issue = {4},
    pages = {043501},
    year = {2023},
    month = {April},
    publisher = {American Physical Society},
    doi = {10.1103/PhysRevAccelBeams.26.043501},
    url = {https://link.aps.org/doi/10.1103/PhysRevAccelBeams.26.043501}
}

@misc{simons2025iotaexperimentprotonpulse,
    title={{IOTA Experiment for Proton Pulse Compression at Extreme Space-Charge}},
    author={Benjamin Simons and Nilanjan Banerjee and Jeffrey Eldred and Vladimir Shiltsev},
    year={2025},
    eprint={2512.03002},
    archivePrefix={arXiv},
    primaryClass={physics.acc-ph},
    url={https://arxiv.org/abs/2512.03002}
}

@article{McDonald:2009zz,
    author = {McDonald, Kirk T. and others},
    editor = {Comyn, Martin and Koscielniak, Shane and Schaa, Volker R. W. and Schmor, Paul W.},
    title = {{The MERIT High-Power Target Experiment at the CERN PS}},
    reportNumber = {FERMILAB-CONF-10-212-APC, PAC09-TU4GRI03},
    journal = {Conf. Proc. C},
    volume = {090504},
    pages = {795--799},
    year = {2009}
}

@misc{RaDIATE,
    url = {https://radiate.fnal.gov/},
    title = {{The RaDIATE Collaboration}}
}

@misc{mu2e,
    url = {https://mu2e.fnal.gov/},
    title = {{Mu2e: muon-to-electron-conversion experiment}}
}

@misc{usmdp,
    url = {https://usmdp.lbl.gov/},
    title = {{U.S. Magnet Development Program}}
}

@article{PhysRevAccelBeams.23.072001,
    title = {Operation of normal-conducting rf cavities in multi-{T}esla magnetic fields for muon ionization cooling: A feasibility demonstration},
    author = {Bowring, D. and Bross, A. and Lane, P. and Leonova, M. and Moretti, A. and Neuffer, D. and Pasquinelli, R. and Peterson, D. and Popovic, M. and Stratakis, D. and Yonehara, K. and Kochemirovskiy, A. and Torun, Y. and Adolphsen, C. and Ge, L. and Haase, A. and Li, Z. and Martin, D. and Chung, M. and Li, D. and Luo, T. and Freemire, B. and Liu, A. and Palmer, M.},
    journal = {Phys. Rev. Accel. Beams},
    volume = {23},
    issue = {7},
    pages = {072001},
    year = {2020},
    month = {July},
    publisher = {American Physical Society},
    doi = {10.1103/PhysRevAccelBeams.23.072001},
    url = {https://link.aps.org/doi/10.1103/PhysRevAccelBeams.23.072001}
}

@misc{osti_1221361,
    author       = {Stratakis, Diktys and Neuffer, David V.},
    title        = {{Utilizing Gas Filled Cavities for the Generation of an Intense Muon Source}},
    url          = {https://www.osti.gov/biblio/1221361},
    place        = {United States},
    organization = {Fermi National Accelerator Laboratory (FNAL), Batavia, IL},
    year         = {2015},
    month        = {May}
}

@misc{candelhigh,
    title={HIGH-FIDELITY INJECTOR MODELING WITH PARALLEL FINITE ELEMENT {3D} ELECTROMAGNETIC {PIC} CODE {Pic3P}},
    author={Candel, A and Kabel, A and Lee, L and Li, Z and Ng, C and Schussman, G and Ko, K},
    url          = {https://proceedings.jacow.org/ICAP2009/papers/we4iopk02.pdf}
}

@misc{ace3p,
    title        = {{ACE3P}},
    url          = {https://confluence.slac.stanford.edu/spaces/AdvComp/pages/240264211/Materials+for+CW23}
}

@techreport{pellemoine2025high,
    title={{High-Power Targetry R\&D Roadmap for High Energy Physics}},
    author={Pellemoine, Frederique and Ganguly, Sudeshna},
    year={2025},
    institution={Fermi National Accelerator Laboratory (FNAL), Batavia, IL (United States)}
}

@misc{Acosta2021,
    author        = {Darin Acosta and Wei Li},
    title         = {{A Muon--Ion Collider at BNL:} The Future {QCD} Frontier and Path to a New Energy Frontier of $\mu^+\mu^-$ Colliders},
    year          = {2021},
    eprint        = {2107.02073},
    archivePrefix = {arXiv},
    primaryClass  = {physics.acc-ph},
    url           = {https://arxiv.org/abs/2107.02073}
}

@misc{Hamada2022,
    author        = {Yu Hamada and others},
    title         = {{$\mu$TRISTAN}},
    year          = {2022},
    eprint        = {2201.06664v2},
    archivePrefix = {arXiv},
    primaryClass  = {physics.acc-ph},
    url           = {https://arxiv.org/abs/2201.06664v2}
}

@inproceedings{latina:hb2023-wea3c1,
    author       = {A. Latina},
    title        = {{The Tracking Code RF-Track and Its Application}},
    booktitle    = {Proc. 68th Adv. Beam Dyn. Workshop High-Intensity High-Brightness Hadron Beams (HB'23)},
    eventdate    = {2023-10-09/2023-10-13},
    pages        = {245--248},
    paper        = {WEA3C1},
    language     = {english},
    venue        = {Geneva, Switzerland},
    series       = {ICFA Advanced Beam Dynamics Workshop on High-Intensity and High-Brightness Hadron Beams},
    number       = {68},
    publisher    = {JACoW Publishing, Geneva, Switzerland},
    month        = {April},
    year         = {2024},
    issn         = {2673-5571},
    isbn         = {978-3-95450-253-0},
    doi          = {10.18429/JACoW-HB2023-WEA3C1},
    url          = {https://jacow.org/hb2023/papers/wea3c1.pdf}
}

@inproceedings{Schulte:1999,
    author = {Schulte, D.},
    title = {{Beam-Beam Simulation with GUINEA-PIG}},
    booktitle = {5th International Computational Accelerator Physics Conference},
    year = {1999},
    url = {https://www.slac.stanford.edu/econf/C980914/papers/F-Th24.pdf}
}

@article{10.1063/1.1818445,
    author = {Neuffer, David},
    title = {``High Frequency'' Buncher and Phase Rotation},
    journal = {AIP Conference Proceedings},
    volume = {721},
    number = {1},
    pages = {407--412},
    year = {2004},
    month = {November},
    issn = {0094-243X},
    doi = {10.1063/1.1818445},
    url = {https://doi.org/10.1063/1.1818445},
    eprint = {https://pubs.aip.org/aip/acp/article-pdf/721/1/407/11404241/407_1_online.pdf}
}

\end{document}